\documentclass[fleqn,usenatbib]{mnras}

\usepackage{newtxtext,newtxmath}

\usepackage[T1]{fontenc}

\usepackage{graphicx}	
\usepackage{amsmath}	
\usepackage{enumitem}

\newcommand{\hi}{H\textsc{i}\ }
\newcommand{\hinospace}{\textrm{H\textsc{i}}}

\newcommand{\Tbar}{\langle T_\hinospace \rangle}
\newcommand{\NFG}{N_\text{FG}}

\newcommand{\size}[2]{{\fontsize{#1}{0}\selectfont#2}}

\newcommand{\fastica}{F\size{7.5}{AST}ICA}
\newcommand{\secref}[1]{\hyperref[#1]{Section~\ref*{#1}}}
\newcommand{\appref}[1]{\hyperref[#1]{Appendix~\ref*{#1}}}

\usepackage[dvipsnames]{xcolor}

\newcommand{\oldred}[1]{\textcolor{black}{#1}}
\newcommand{\red}[1]{\textcolor{black}{#1}}

\setlist[enumerate,1]{leftmargin=*}

\title[21cm foregrounds: cleaning and mitigation]{21cm foregrounds and polarisation leakage: a user’s guide on cleaning and mitigation strategies}

\author[S. Cunnington et al.]{
Steven Cunnington,$^{1}$\thanks{E-mail: s.cunnington@qmul.ac.uk}
Melis O. Irfan,$^{2,3}$
Isabella P. Carucci,$^{3}$
Alkistis Pourtsidou,$^{1,2}$
\newauthor
J\'{e}r\^{o}me Bobin$^{3}$
\\
$^{1}$School of Physics and Astronomy, Queen Mary University of London, Mile End Road, London E1 4NS, UK\\
$^{2}$Department of Physics \& Astronomy, University of the Western Cape, Cape Town 7535, South Africa\\
$^{3}$AIM, CEA, CNRS, Université Paris-Saclay, Université Paris Diderot, Sorbonne Paris Cité, F-91191 Gif-sur-Yvette, France
}

\date{Accepted XXX. Received YYY; in original form ZZZ}

\pubyear{2020}

\begin{document}
\label{firstpage}
\pagerange{\pageref{firstpage}--\pageref{lastpage}}
\maketitle

\begin{abstract}
The success of \hi intensity mapping is largely dependent on how well 21cm foreground contamination can be controlled. In order to progress our understanding further, we present a range of simulated foreground data from \oldred{two} different $\sim3000$\,deg$^2$ sky regions, with \oldred{varying effects} from polarisation leakage. Combining these with cosmological \hi simulations creates a range of intensity mapping test cases that require different foreground treatments. This allows us to conduct the most generalised study to date into 21cm foregrounds and their cleaning techniques for the post-reionisation era. We first provide a pedagogical review of the most commonly used blind foreground removal techniques (PCA/SVD, \fastica, GMCA). We also trial a non-blind parametric fitting technique and discuss potential hybridization of methods. We highlight the similarities and differences in these techniques finding that the blind methods produce near equivalent results, and we explain the fundamental reasons for this. 
Our results demonstrate that polarised foreground residuals should be generally subdominant to \hi on small scales ($k\gtrsim0.1\,h\,\text{Mpc}^{-1}$). However, on larger scales, results are more \oldred{case-dependent}. In some cases, aggressive cleans severely damp \hi power but still leave dominant foreground residuals. \oldred{We find a changing polarisation fraction has little impact on results within a realistic range (0.5\% - 2\%), however a higher level of Faraday rotation does require more aggressive cleaning.} We also demonstrate the gain from cross-correlations with optical galaxy surveys, where extreme levels of residual foregrounds can be circumvented. However, these residuals still contribute to errors and we discuss the optimal balance between over- and under-cleaning. 

\end{abstract}

\begin{keywords}
cosmology: large scale structure of Universe -- cosmology: observations -- radio lines: general -- methods: data analysis -- methods: statistical
\end{keywords}



\section{Introduction}

Mapping the cosmic neutral hydrogen (\hinospace) from the post-reionisation era is as an excellent way to probe the large-scale structure of the Universe. By mapping the redshifted 21cm signal from \hi residing within galaxies, the underlying 3-dimensional matter density can be inferred and cosmological information can be extracted, in a similar fashion to optical galaxy surveys. A novel technique allowing to do this is intensity mapping \citep{Bharadwaj:2000av,Battye:2004re,Chang:2007xk}.

In this work we focus on so-called \textit{single-dish} intensity mapping \citep{batps}, which uses the auto-correlation data of a telescope array (e.g. the Square Kilometer Array (SKA) -- \citet{Bacon:2018dui}), as opposed to the more traditional interferometric mode of operation. Unlike a conventional spectroscopic galaxy survey that has to resolve galaxies and conduct spectroscopy to infer a redshift with sufficient precision, intensity mapping does not resolve galaxies but records the diffuse, unresolved \hinospace. This has the advantages of being able to rapidly observe very large volumes of the Universe, and is not as susceptible to high levels of shot-noise. The resulting maps have a relatively low-angular resolution due to the radio telescope beam, which is related to the dish diameter for single-dish observations. This damps the \hi power spectrum for modes perpendicular to the line-of-sight but despite this, many large cosmological scales can still be resolved. Furthermore, the spectroscopic resolution in these radio observations is excellent and thus modes can in principle be resolved to very small scales along the line-of-sight \citep{Villaescusa-Navarro:2016kbz}.

There are unique challenges to \hi intensity mapping which conventional galaxy surveys largely avoid. Whilst intensity mapping is unlikely to be limited by shot noise, there is instrumental (thermal) noise. Assuming enough observation time, and a well controlled system temperature, this noise should be sub-dominant relative to the \hi signal and well approximated as Gaussian white-noise. As noted in \citet{Harper:2017gln}, complications from other systematics such as $1/f$ noise  pose a more complex challenge. However, analysis of recent \hi intensity mapping data from MeerKAT suggest this should be a controllable systematic \citep{Li:2020bcr}. A further issue is contamination from human-made Radio Frequency Interference (RFI) such as global navigation satellites \citep{Harper:2018ncl}.

Another major challenge, and the focus of this paper, is foreground contamination from astrophysical sources and how they interact with the telescope. The main source of 21cm foreground signals comes from \textit{Galactic synchrotron} (sourced by cosmic-ray electrons accelerated by the Galactic magnetic field), \textit{free-free emission} (sourced by free electrons scattering off ions largely within our Galaxy but weaker emission can also come from extragalactic sources), and \textit{point-sources} (extragalactic objects emitting strong radio signals e.g. AGNs).

Some of these foregrounds can be orders of magnitude more dominant than the \hi signal, but their spectrum evolves slowly through frequency. This is in contrast to cosmic \hinospace, which varies with redshift and thus oscillates to a near-Gaussian approximation with frequency. The fact that the raw-foregrounds are smooth continuums through frequency means they can in principle be removed with modelling or source separation \citep{Liu:2011hh,Wolz:2013wna,Shaw:2014khi,Alonso:2014dhk}. However, large-scale foreground signals typically have some degeneracy with the cosmological \hinospace, and require some form of treatment in order not to bias power spectra measurements and cosmological parameter estimation \citep{Wolz:2013wna, Cunnington:2020wdu,Soares:2020zaq}.

In reality the challenge of separating the \hi signal from the foregrounds becomes even more complicated by the foreground's response to the instrument. Unless instrumental effects from spectral response and chromaticity from the beam are controlled, the spectral smoothness of the foregrounds can be degraded. The most potentially concerning instrumental effect is from polarisation leakage \citep{Jelic:2008jg, Jelic:2010vg, Moore:2013ip}. Cosmological \hi is unpolarised and thus attempts are made to sufficiently calibrate telescopes to avoid polarised signals \citep{Liao:2016jto}. However, a sufficient level of calibration is not guaranteed and even a small amount of polarised synchrotron leaking into the observational data can dominate the \hi signal. Furthermore, the Faraday rotation that interferes with the polarisation state is expected to be frequency-dependent, which means these leaked signals will not have such a smooth spectrum and will be harder to single out \citep{Carucci:2020enz}.

Previous investigations into foreground cleaning generally involve introducing a single set of foreground simulations which cleaning techniques can then be tuned to. However, these rarely include instrumental response effects such as polarisation leakage (although there are some exceptions e.g. \citet{Shaw:2014khi,Carucci:2020enz}). These idealised simulated foregrounds require much less aggressive cleans than what is usually needed in real data analyses from pathfinder intensity mapping experiments \citep{Masui:2012zc,Switzer:2013ewa,Wolz:2015lwa,Anderson:2017ert,Wolz:2021ofa}. 

In this work, we add an extra layer of complication. Whilst we cannot yet provide full end-to-end simulations that directly mimic a realistic experiment,  we are able to present results where it is necessary to use aggressive foreground cleans akin to those employed in real data. By using \oldred{two} different sky regions, \oldred{with varying polarisation leakage effects}, we create a variety of cases in which different levels of foreground cleaning are required and different problems arise. We introduce and apply a range of different foreground cleaning methods and compare the results. Since we are dealing with simulations, we have full control over the data and provide analysis into problems concerning damping of \hi power as well as the biases and errors introduced from foreground residuals.

This paper is structured as follows. In \secref{sec:FGsims} we present our foreground simulations, \oldred{the sky regions we consider and the different cases of polarisation leakage}. In \secref{sec:CosmoSim} we present our method for producing the underlying cosmological \hi intensity maps (and the accompanying optical galaxy data for cross-correlations) that we aim to recover. We provide a generalised review of foreground cleaning methods in \secref{sec:FGremovalSec} and identify the exact methods we apply to our simulated data. We then present our results in \secref{sec:Results} and conclude in \secref{sec:Conclusion}.

\section{Foreground Simulations}\label{sec:FGsims}

\begin{figure}
    \centering
    \includegraphics[width=\columnwidth]{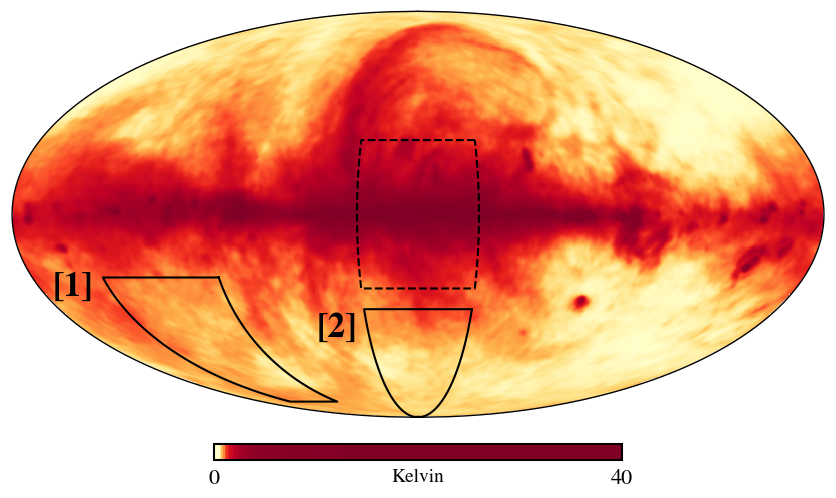}
    \includegraphics[width=\columnwidth]{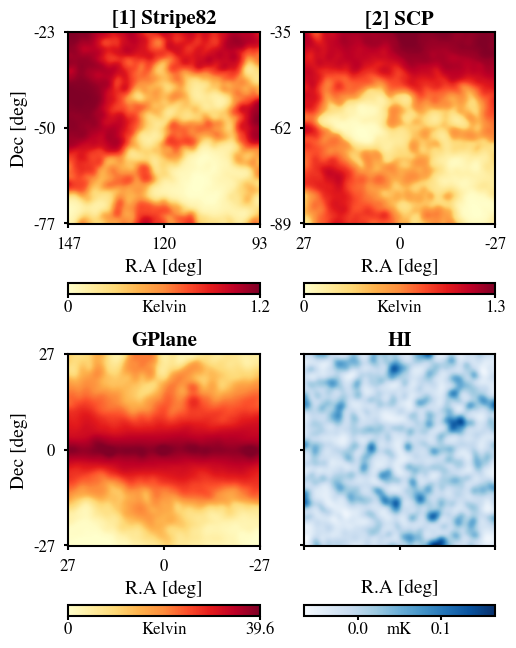}
    \caption{(\textit{Top} map): Full sky simulated synchrotron, free-free emission and point sources with an 80 arcmin resolution. The labelled black boarders indicate the position of \oldred{both sky regions we investigate. \textit{(Middle} two maps): Both regions interpolated over a $256^2$ pixel grid. For comparison, the \textit{bottom-left} map shows the Galactic plane (position marked by the \textit{dashed} outline in the top map) which is orders of magnitude higher in emission. \textit{Bottom-right} is the cosmological \hi (introduced in \secref{sec:HIsims}) which we overlay onto both regions [1] and [2] and attempt to recover.} All maps are at $1050\,\text{MHz}$ and include the effects from a $1.67\,\text{deg}$ telescope beam.}
    \label{fig:regions}
\end{figure}

We begin by identifying \oldred{two} different regions on the sky. We desire each of our regions to have the same size which is dictated by the size of our 1\,$h^{-3}$Gpc$^3$ \hi simulation box (to be described in detail in \secref{sec:CosmoSim}). At the central redshift of this simulation ($z=0.39$) these dimensions are equivalent to a sky size of $54.1 \times 54.1 \sim 2972\,$deg$^2$. This is similar to the sky area proposed in MeerKLASS \citep{Santos:2017qgq}, a wide area survey using the MeerKAT telescope, which is the pathfinder for the Square Kilometre Array (SKA)\footnote{\href{https://www.skatelescope.org/}{skatelescope.org}} \citep{Bacon:2018dui}. We choose a frequency range of $899 - 1184$\,MHz ($0.2<z<0.58$), again consistent with a MeerKAT-like observation performed in the L-band. The sky regions we investigate are:\newline
\\
\noindent \textbf{[1] Stripe 82}:\\
A small, 300\,deg$^2$ field imaged numerous times by galaxy surveys. To ensure consistency in sky sizes, this region is a 2927\,deg$^{2}$ patch centred on the Stripe 82 field. \newline
\\
\noindent \textbf{[2] South Celestial Pole (SCP)}:\\
Low declination region away from Galactic Plane where combined emission from all foregrounds is expected to be low. \newline
\\
\noindent These regions are outlined in \autoref{fig:regions} on the full-sky (top-map) and individually shown in the flattened maps below. We assume the maps are sufficiently small in size that they can be projected onto a Cartesian grid with minimal distortion. For each of these regions we simulate effects from polarisation leakage, which we discuss further in \secref{sec:PolarisatitonLeak}.

The total observed temperature data are a combination of the cosmological \hi signal, the foregrounds and instrumental noise, all binned into pixels whose position is defined by $\boldsymbol{\theta}$ at each frequency channel $\nu$
\begin{equation}\label{eq:DecompSignal}
	\delta T_\text{obs}(\nu,\boldsymbol{\theta}) = \delta T_\hinospace(\nu,\boldsymbol{\theta}) + \delta T_\text{FG}(\nu,\boldsymbol{\theta}) + \delta T_\text{noise}(\nu,\boldsymbol{\theta})\,.
\end{equation}
In this work, we will vary $\delta T_\text{FG}$ between each region whilst keeping the other components fixed. The foreground signal can be further decomposed into the contributions from the different sources of foregrounds i.e. Galactic synchrotron emission, Galactic free-free emission, extragalactic point sources and polarisation leakage :
\begin{equation}\label{eq:FGdecomp}
	\delta T_\text{FG} = \delta T_\text{syn} + \delta T_\text{free} + \delta T_\text{point} + \delta T_\text{pol}\, .
\end{equation}
We introduce our simulation approach for each of these components in this section (except the \hi contribution which is discussed in \secref{sec:CosmoSim}). A full-sky realisation of $\delta T_\text{FG}$ is openly available in \citet{FGsim}.

We chose the frequency range $899 - 1184$\,MHz and separate this range into 285 measurement bands. The effective resolution is determined by the beam size of the instrument which is dependent on the frequency $\nu$ and therefore each channel is smoothed by a different amount. However, foreground removal algorithms (discussed in \secref{sec:FGremovalSec}) perform better on data with a common resolution. We therefore smooth the intensity maps to a constant beam size given by the minimum frequency $\nu_\text{min}=899\,\text{MHz}$. \oldred{The full-width-half-maximum of the beam is given by}
\begin{equation}\label{eq:BeamEq}
    \oldred{\theta_{\mathrm{FWHM}}=\frac{1.18c}{\nu D_\text{dish}}.}
\end{equation}
where $c$ is the speed of light and we assume a dish size of \oldred{$D_\text{dish}=13.5\,\text{m}$, which is the size of MeerKAT's dishes and is approximately equivalent to the SKA-MID's dishes too. The factor of $1.18$ can vary depending on the beam pattern but we chose this value to be consistent with the recent MeerKAT investigations in \citet{Matshawule:2020fjz}. From \autoref{eq:BeamEq}, we therefore get an effective resolution for our maps of $\theta_{\mathrm{FWHM}}=1.67\,\text{deg}$}. We chose to create our simulations at $N_\text{side} = 2048$ and then to interpolate our $54.1 \times 54.1$ patches onto $256 \times 256$ pixel arrays.

We make use of the  Planck Legacy Archive\footnote{\href{http://pla.esac.esa.int/pla}{pla.esac.esa.int/pla}} FFP10 simulations within our simulations and, as they are given in $T_{\rm{CMB}}$, the following conversion to the Rayleigh-Jeans regime is used:
\begin{equation}
    T_{\rm{RJ}} = \frac{x^{2}  e^{x}}{(e^{x} - 1)^2} T_{\rm{CMB}}, 
 \end{equation}
where $x = h \, \nu/k_\text{B} \, T_{\rm{CMB}}$, with $h$ the Planck constant and $k_\text{B}$ the Boltzmann constant.

\subsection{Simulated Synchrotron Emission}

We use the FFP10 simulations of synchrotron emission at 217 and 353\,GHz for our purposes as these maps are provided at $N_\text{side}=2048$. These maps are formed from the source-subtracted and destriped 0.408\,GHz map \citep{newHas}. Despite the 0.408\,GHz survey data having a resolution of 56 arcmin, \citet{newHas} provide a $N_\text{side}=2048$ version of the data by filling in the higher resolution detail with a Gaussian random field. 

These 217 and 353\,GHz synchrotron maps can be used to determine the synchrotron spectral index map at $N_\text{side}=2048$. The spectral index map used by FFP10 is the 
`Model 4' synchrotron spectral index map of \citet{mamd08}, which has a resolution of $\sim 5$ degrees. This map was formed from 0.408\,GHz intensity data and 23\,GHz polarisation data. However, as we are simply trying to determine the accuracy of our foreground mitigation strategies, the accuracy of the synchrotron spectral index map does not come into consideration.

We will however, need a higher resolution view of the synchrotron spectral index than 5 degrees and so we also choose to fill in the higher resolution detail with a Gaussian random field. Taking the synchrotron multipole scaling relation from \citet{santos05}, our $N_\text{side}=2048$ synchrotron spectral index map is constructed as
\begin{equation}
    \beta_\text{sy} = \beta_\text{model4} + \beta_\text{ss}\,,
\end{equation}
where
\begin{equation}
    C_{\ell}^{\beta_\text{ss}} =  7 \times 10^{-6} \left( \frac{1000}{\ell} \right)^{2.4}  \left(\frac{\nu_{r}^{2}}{\nu_{1} \nu_{2}}\right)^{2.8} {\rm{exp}}\left(\frac{- {\rm{log}}(\nu_{1} / \nu_{2})^{2}}{2 \times 4^{2}} \right)\,,
\end{equation}
where $\nu_{r}$ is 130\,MHz, $\nu_{1}$ is 580\,MHz and $\nu_{2}$ is 1000\,MHz. Our Gaussian random field is identical to that found in \citet{santos05} with the exception of the amplitude, which we alter to suit the magnitude of the synchrotron spectral index as opposed to the emission amplitude. \oldred{We then smooth $\beta_\text{sy}$ to 1.67 degrees in order to match the desired resolution of our total simulation maps.}

 \subsection{Simulated Free-Free Emission}
 
We take our simulated free-free amplitude ($a_\text{ff}$) from the FFP10 217\,GHz free-free simulation at $N_\text{side}=2048$. This map is a composite of the \citet{cliveff}  free-free template and the WMAP MEM free-free templates; the details of which can be found in \citep{mamd08}. Our free-free emission is modelled by a power law
\begin{equation}
    T_\text{ff}(\nu, \boldsymbol{\theta}) = a_\text{ff}(\boldsymbol{\theta}) \left(\frac{\nu}{\nu_{0}} \right)^{\beta_\text{ff}}\,,
\end{equation}
where the free-free spectral index is $\beta_\text{ff}=-2.13$ and constant across all map pixels.

\subsection{Simulated Point Sources}
 
We use the empirical model of \citet{batps}, which fits a polynomial to a selection of radio source counts at 1.4\,GHz. The specific details of assembling this model of the Poisson and clustering contributions at 1.4\,GHz can be found in \citet{thesis}. Following the method of \cite{oliv}, we then scale the 1.4\,GHz point source map to our frequencies using a power law where the spectral index varies following a Gaussian distribution centred at -2.7, with a standard deviation of 0.2.
 
\citet{batps} expect point sources over 10\,mJy to be bright enough to be identified within the National Radio Astronomy Observatory Very Large Array Sky Survey \citep{vlass} and so, removed. In this work we consider a 100\,mJy upper bound on source extraction.

\subsection{Simulated Polarisation Leakage}\label{sec:PolarisatitonLeak}

The magnetic fields within our Galaxy's interstellar medium can cause Faraday rotation effects which change the polarisation angles of light. If this were a consistent effect, it would not be hugely problematic for foreground classification. However, Faraday rotation is a frequency-dependent effect as demonstrated by \citet{Jelic:2010vg,Moore:2013ip}. If any spectrally fluctuating polarisation intensity is leaked into the total intensity it would be difficult to subtract without large loss to the unpolarised \hi cosmological modes. Depending on both the instrument and the data reduction scheme implemented, there will be some percentage of leakage of Stokes Q and U synchrotron emission into Stokes I. Faraday rotation alters the true polarisation angle of the Stokes Q/U signal such that this leakage will not remain constant across all the observational channels.

We simulate this instrumental effect with the use of the \texttt{CRIME}\footnote{\href{http://intensitymapping.physics.ox.ac.uk/CRIME.html}{intensitymapping.physics.ox.ac.uk/CRIME.html}} software, which provides maps of Stokes Q emission at each frequency \oldred{with a choice for the polarisation leakage fraction, typically between 0.5 and 1\% \citep{Liao:2016jto}}. Further observational data analysis would be needed to constrain a reasonable choice for this fraction \oldred{so we therefore explore a range of polarisation leakage fractions}. Details for the rotation calculation of the Stokes Q synchrotron emission from Faraday depth measurements \citep{opp} are given in \citet{crime}. 

\oldred{We highlight that the polarisation leakage model we use has many limitations. For example, it assumes that all the leakage comes from just the Q Stokes component and not U, reducing possible mixtures; it uses rotation measures of extra-galactic sources \citep{opp}, corresponding to the maximum rotation from a single source; it does not account for multiple Faraday rotating components along a single
line-of-sight. These issues could give rise to a more complex structure for this systematic. However, at present, the community lacks a better model than what has been proposed by \citet{crime}. \citet{Shaw:2014khi} also assembled a polarisation leakage model using the rotation measure map of \citet{opp}, however showing a qualitatively different structure in pixel-space. The two polarisation leakage models differ in their choices of polarisation fraction magnitude and correlation length in frequency space. Both models are valid within the current knowledge of intensity mapping polarisation leakage.} \oldred{The lack of smoothness in frequency is what makes polarisation leakage a challenging component to separate. Therefore, to be conservative, we decide to make use of our simulated polarisation leakage both in its {\it milder} regions and the most troublesome one, where a higher Faraday rotation causes increased decorrelation (or a lack of smoothness). For each of the two sky regions we look at, we therefore provide two polarisation leakage cases; one predicted by the \texttt{CRIME} model for that region, and a second from the \texttt{CRIME} output for the Galactic plane (see dashed boarder in \autoref{fig:regions}). The latter stronger model of polarisation leakage we refer to as our high-Faraday rotation (FR) case.}

\begin{figure}
    \centering
    \includegraphics[width=\columnwidth]{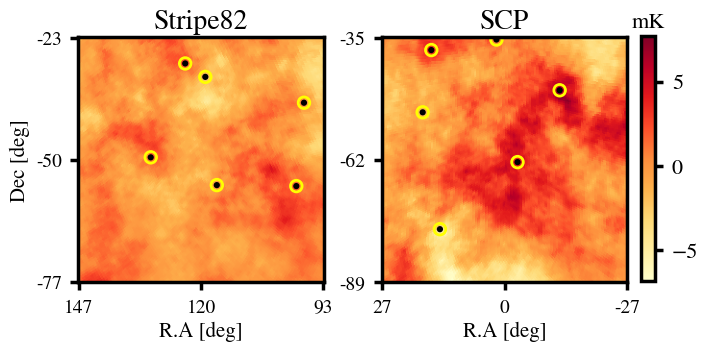}
    \caption{\oldred{Polarisation leakage maps from \texttt{CRIME} for the both regions averaged along the line-of-sight. The \textit{yellow-black} markers indicate the  positions of the random pixels used for the $\delta T_\text{pol}$ line-of-sight spectra in \autoref{fig:FGSpectra}.}}
    \label{fig:PolMaps}
    \centering
    \includegraphics[width=\columnwidth]{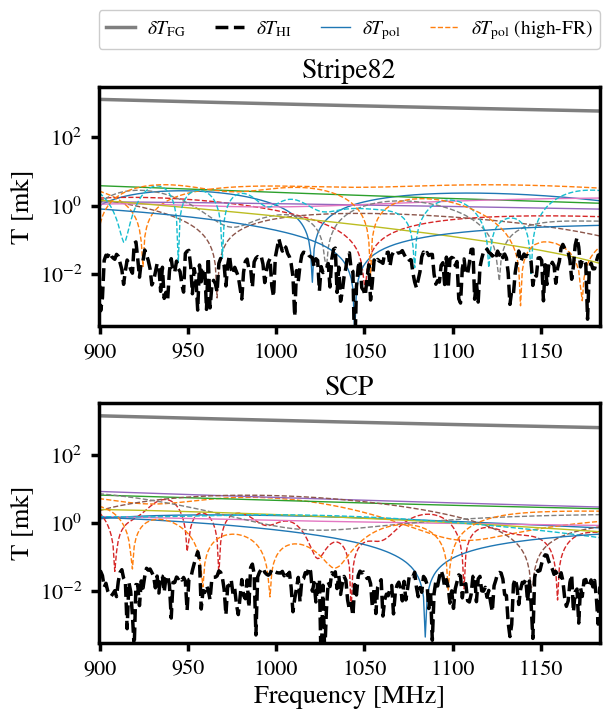}
    \caption{\oldred{Spectra for the different components of the the observed signal. \textit{Grey-solid} line shows the total foreground component (\autoref{eq:FGdecomp}) for one random line-of-sight and \textit{black-dashed} line shows \hinospace-only. Smoothness of the foregrounds and their large amplitude relative to the \hi is clear and are the distinguishing features utilised in a foreground clean. The \textit{solid-coloured-thin} lines show six random lines-of-sight for the polarised contribution. The positions of these lines-of-sight in angular space are marked in \autoref{fig:PolMaps}. The polarised components cause a decoherence to the spectra creating difficulties in a foreground clean. We also show the polarised data taken from the Galactic plane (\textit{dashed-coloured-thin} lines) which we refer to as a high Faraday rotation (FR) case.}}
    \label{fig:FGSpectra}
\end{figure}

\oldred{\autoref{fig:PolMaps} shows the output maps for the polarisation leakage component from \texttt{CRIME} for our two regions. The SCP regions contains slightly higher amplitudes than Stripe82 however, as discussed above, the main problem that polarisation leakage introduces is frequency decoherence. In \autoref{fig:FGSpectra} we show the spectra for some random lines-of-sight in the polarised maps (coloured-solid lines). We find the Stripe82 region contains more oscillating spectra in this polarisation model and we therefore expect this to be the more troublesome regions in our foreground tests. We also include in \autoref{fig:FGSpectra} the spectra from random lines-of-sight in the high Faraday rotation case we use (data from the Galactic plane), shown as coloured-dashed lines. We can see for this case the oscillations are more extreme and when these are incorporated into the total observed signal they will cause more frequency decoherence and represent the most challenging case for foreground cleaning.}

\subsection{Simulated Noise}\label{sec:NoiseSim}

In this work, only Gaussian noise is considered, with a zero mean and standard deviation of
\begin{equation}\label{eq:NoiseEq}
    \sigma (\nu) =  T_{\rm{sys}}(\nu) \left( \delta_{\nu} \, t_{\rm{tot}} \, \frac{\Omega_\text{p}}{\Omega_\text{a}} \, N_{\rm{dish}} \right)^{-1/2},
\end{equation}
where $\delta_{\nu}$ is the width of each frequency band (Hz),  $t_{\rm{tot}} $ is the total survey time (s), $N_{\rm{dish}}$ is the number of dishes and $\Omega_\text{p/a}$ are the pixel and survey solid angle, respectively \citep{crime}. For the pixel solid angle only the beam FWHM expressed in radians is required
\begin{equation}
    \Omega_\text{p} = 1.13 \, \theta_{\rm{FWHM}}^{2}\,,
\end{equation}
while for the survey solid angle the fraction of the sky covered is needed. If the angular area of the observed sky ($A_\text{sky}$) is given in square degrees, we have 
\begin{equation}
    \Omega_\text{a} = 4 \pi \frac{A_\text{sky}}{41253}.
\end{equation}
The system temperature in each band ($T_{\rm{sys}}(\nu)$) is a combination of the receiver noise temperature ($T_{\rm{rec}}$) and the sky temperature \citep{skytemp}:
\begin{equation}
    T_{\rm{sys}}(\nu) = 1.1 \times 60 \left(\frac{300}{\nu [{\rm MHz}]}\right)^{2.55} + T_{\rm{rec}}.
\end{equation}
The specific receiver and survey properties used here are based on a MeerKLASS-like survey and are summarised in \autoref{tab:table3}.
 
\begin{table}
    \caption{The assumed receiver and survey properties for observation with bandwidth $899 < \nu < 1184\,\text{MHz}$ ($0.2<z<0.58$).}
    \label{tab:table3}
    \centering
    \begin{tabular}{lccccc}
    \hline
    {\bf{Quantity}:} & $\delta_{\nu}$ & $t_{\rm{tot}}$ & $N_{\rm{dish}}$ & $T_{\rm{rec}}$ & $A_\text{sky}$ \\
    {\bf{Value}:} & 1\,MHz & 1000 hrs & 64 & 25\,K & 2927 deg$^{2}$ \\
    \hline
    \end{tabular}
\end{table}

\section{Cosmological Simulations}\label{sec:CosmoSim}

We use the same simulated cosmological \hi signal data for each sky region. Specifically, we make use of the \textsc{MultiDark-Planck} cosmological $N$-body simulation \citep{Klypin:2014kpa}, which evolved $3840^3$ dark-matter particles in a $1000^3\,h^{-3}\,\text{Mpc}^3$ volume with the adopted cosmology complying with \textsc{Planck15} \citep{Ade:2015xua}. The cosmological parameters used are therefore $\Omega_\text{M} = 0.307$, $\Omega_\text{b} = 0.048$, $\Omega_\Lambda = 0.693$, $\sigma_8 = 0.823$, $n_\text{s} = 0.96$ and Hubble parameter $h=0.678$. This data has been processed into the \textsc{MultiDark-Galaxies} data \citep{Knebe:2017eei}, which are galaxy catalogues publicly available from the Skies \& Universes web page\footnote{\href{http://www.skiesanduniverses.org/page/page-3/page-22/}{www.skiesanduniverses.org}}. It is from these catalogues that we build the simulated \hi intensity maps and an overlapping map of resolved optical galaxies.

Each snapshot from the \textsc{MultiDark-Galaxies} simulation represents a different redshift and evolved state of the cosmological density field and the galaxies therein. We opt to use the catalogues at $z=0.39$ and take this as the effective redshift ($z_\text{eff}$) for our data. This is analogous to real surveys assuming a central effective redshift provided that the width of the bin is small enough so that cosmological quantities can be assumed constant within it. 

We still need to assume some redshift range however, since we require a frequency range from which to produce the foregrounds. We therefore assume our data has redshift range of $0.2<z<0.58$, which for the \hi intensity maps with $\nu = 1420\,\text{MHz}/(1+z)$, will convert to a frequency range of $899<\nu<1184\,\text{MHz}$. This frequency range is probed with the L-band from the MeerKAT telescope and is thus representative of a near-term intensity mapping survey \citep{Santos:2017qgq}.

The \textsc{MultiDark} data we use are for a Cartesian box with galaxy coordinates in physical distances. We thus work in this Cartesian regime throughout this investigation. This is common practice in large-scale structure surveys, where either a small enough sky is surveyed that a flat-sky approximation is valid, or where curved sky effects  are accounted for \citep{Castorina:2017inr,Blake:2018tou}. At the effective redshift $z_\text{eff}=0.39$, the redshift range of $0.2<z<0.58$ we assume for our data converts to a physical distance of $925\,h^{-1}\,\text{Mpc}$. We therefore trim the \textsc{MultiDark} data cube to this distance along one dimension, keeping the others the same. This results in a data cube with physical size $L_\text{x},L_\text{y},L_\text{z} = 1000,1000,925\,h^{-1}\,\text{Mpc}$ where we use the convention that x and y are the angular dimensions perpendicular to the line-of-sight and z is parallel to the line-of-sight. We use the plane-parallel approximation throughout. The data cube is gridded into volume-pixels (voxels), with $n_\text{x}=n_\text{y}=256$ along the angular dimensions and $n_\text{z}=285$ along the radial dimension. The choice of radial binning allows the $899<\nu<1184\,\text{MHz}$ frequency range we assume to have a frequency resolution of $\delta\nu=1\,\text{MHz}$.
As we have already mentioned, the approximate sky coverage of our data is just under $3000\,\text{deg}^2$, which is fairly representative of proposed intensity mapping surveys like MeerKLASS \citep{Santos:2017qgq}.

\subsection{\hi Intensity Maps}\label{sec:HIsims}

To produce the intensity maps from the \textsc{MultiDark} data we utilise the catalogue produced from applying the SAGE \citep{Croton:2016etl} semi-analytical model to the data. We summarise our method below for how we produce the intensity maps from the \textsc{MultiDark-SAGE} catalogue. For a more complete description of this process, we refer the reader to \citet{Cunnington:2020mnn}, where an identical methodology was employed. The \textsc{MultiDark-SAGE} catalogue is fully outlined in \citep{Knebe:2017eei}.

Firstly the cold gass mass for each galaxy is converted into a \hi mass, which is then binned into the relevant voxel according to the galaxy's coordinates. This gridded \hi mass is then converted to a \hi brightness temperature $T_\hinospace(\boldsymbol{x})$. Since intensity mapping surveys will detect signal down to the very faintest of emitters, it is common in simulations to rescale the $T_\hinospace$ temperature of the field up to a realistic (expected) value. This is required because simulations have finite capabilities and often do not resolve halos down to masses of $\sim10^8\,h^{-1}M_\odot$ where \hi is still predicted to reside \citep{Villaescusa-Navarro:2018vsg,Spinelli:2019smg}. To determine this value we utilise the results of the GBT-WiggleZ cross-correlation analysis \citep{Masui:2012zc}, where it was found that the \hi abundance is $\Omega_\hinospace b_\hinospace r=[4.3 \pm 1.1] \times 10^{-4}$, and assume it is constant with redshift. We also take the cross-correlation coefficient to be $r=1$ and use a \hi bias fit from \citet{Villaescusa-Navarro:2018vsg}. For our effective redshift this equates to $b_\hinospace(z_\text{eff})=1.105$.

Lastly, in order to emulate the effects from the radio telescope beam we smooth each channel using $\theta_{\mathrm{FWHM}}=1.67\,\text{deg}$ (as discussed in \secref{sec:FGsims}). The observable field for intensity mapping is the over-temperature field defined as 
\begin{equation}
	\delta T_\hinospace (z) = T_\hinospace(z) - \Tbar = \Tbar \, b_\hinospace(z)\, \delta_\text{M}(z) \, ,
\end{equation}
where $\delta_\text{M}(z)$ is the underlying matter density. \oldred{We have shown an example \hi intensity map in the bottom-right panel of \autoref{fig:regions} for one frequency channel at $1050\,\text{MHz}$.}

\subsection{Overlapping Optical Galaxy Data}\label{OpticalSimsSec}

We also utilise the \textsc{MultiDark-Galaxies} for creating an overlapping optical spectroscopic catalogue, which we will use for investigating cross-correlating techniques between \hi intensity maps and optical galaxy data. For this purpose we use the SAG \citep{Cora:2006mn} semi-analytic model; that is because this catalogue has magnitude outputs for each of the SDSS \textit{ugriz} broad bands, which can be utilised to construct a realistic optical galaxy data set.

Whilst the \textsc{MultiDark-SAG} catalogue does also possess the cold gas mass outputs, and therefore could have also been used to produce the intensity maps, it has fewer galaxies ($\sim3.8\times 10^7$) compared with the \textsc{MultiDark-SAGE} catalogue ($\sim7\times 10^7$) at the snapshot redshift of $z=0.39$. We prefer using \textsc{MultiDark-SAGE} for the intensity maps because it has a higher number of galaxies. Since both SAGE and SAG catalogues are generated from the same underlying \textsc{MultiDark} simulated density field, they should still produce sufficient cross-correlation signals. The \textsc{MultiDark-SAG} catalogue is fully outlined in \citep{Knebe:2017eei}.

Optical galaxy surveys generically operate by constructing a catalogue of resolved galaxies whose luminosity is above some threshold determined by the telescope's sensitivity. As a rather crude emulation of this method, which is sufficient for this investigation, we use the sum of the magnitudes from the five SDSS \textit{ugriz} bands and select the highest total magnitudes from the simulation until a target $N(z)$ redshift distribution is achieved. Following \citet{Mandelbaum:2009ck}, we construct a realistic target distribution by assuming a double Gaussian where 77.6\% of the galaxies are in the first Gaussian and the remaining 22.4\% are in the second Gaussian. The Gaussian's are centred at $\langle z \rangle = 0.595$ and $\langle z \rangle = 0.558$ with standard deviations of $\sigma_z=0.236$ and $\sigma_z=0.112$ respectively. Since we are simulating a spectroscopic redshift galaxy sample, we assume all redshifts have been measured correctly to the precision required for correct binning into our Cartesian grid. Imposing the redshift bin limits for our simulated survey of $z_{\min}=0.2$ and $z_{\max}=0.58$ provides the redshift distribution. We then finally stipulate that $2\times 10^6$ galaxies will be detected in the optical survey. The over-density field for the optical galaxies is given as 
\begin{equation}
	\delta_\text{g}(z) = \frac{n_\text{g}(z) - \langle n_\text{g}\rangle}{\langle n_\text{g}\rangle} = b_\text{g}(z)\,\delta_\text{M}(z) \, ,
\end{equation}
where $b_\text{g}$ is the linear bias for the optical galaxy field. For these simulated galaxy maps and the simulated \hi intensity maps (presented in \secref{sec:HIsims}) we checked that both measured power spectra, and their cross-correlation, are modelled well by commonly used anisotropic redshift space clustering models (see e.g. \citet{Soares:2020zaq}), thus validating their use as our underlying cosmological data.

\section{Methods for Foreground Cleaning}\label{sec:FGremovalSec}

Here we discuss some of the most popular and well studied approaches to 21cm foreground cleaning. Our focus in this work is on single-dish observations, in the context of cosmological analysis and we are therefore ultimately trying to optimise a power spectrum measurement. All foreground removal methods aim to utilise the fact that the foreground contributions are slowly varying with frequency (unlike cosmological \hinospace) and are orders of magnitude more dominant than the \hinospace. Thus, the general approach is identifying a set of smooth functions that represent the dominant foreground contributions and subtracting these from the data to leave the cosmic \hi signal. The method for estimating this set of smooth functions is largely where the techniques diverge into the wide library of foreground removal options available today (see e.g. \citet{Liu:2019awk} for a more detailed summary).  

Blind component separation methods dominate the literature concerning foreground removal techniques, and we also use them in our analysis. Blind separation means little input information is needed and the process exploits the fact that relatively few dominant uncorrelated (or statistically independent) (or sparse) sources should contain the majority of the foreground emission in the observed signal. The advantage of such an approach is that it does not require a detailed understanding of the foreground signals, e.g. their precise amplitude through frequency, and how they respond to instrumental systematics. Given that we are a long way from fully understanding sky emission at the $\sim$21cm wavelengths and that the intensity mapping technique is still in its infancy (meaning instrumental response and systematics are poorly understood), it is sensible for blind methods to be the preferred choice \citep{Masui:2012zc,Wolz:2015lwa,Anderson:2017ert}.

The raw observed sky signal in intensity mapping\footnote{Neglecting contributions from more complex systematics.} can be decomposed into contributions from the cosmological \hinospace, the foregrounds, and the thermal noise from the instrument (as in \autoref{eq:DecompSignal}). These observed data can be represented by a matrix $\mathbf{X}_\text{obs}$ with dimensions $N_\nu \times N_\theta$ where $N_\nu$ is the number of frequency channels along the line-of sight and $N_\theta$ the number of pixels. In this approach the 2D ($N^\text{ra}_\theta, N^\text{dec}_\theta$) angular pixel space is turned into a $N_\theta = N^\text{ra}_\theta \times N^\text{dec}_\theta$ long 1D vector to make the foreground cleaning formalism more concise.

We make the assumption that the data matrix $\mathbf{X}_\text{obs}$ can be represented as a linear system
\begin{equation}\label{eq:FGlinearSys}
	\mathbf{X}_\text{obs} = \mathbf{\hat{A}}  \mathbf{S} + \mathbf{R} \, ,
\end{equation}
where $\mathbf{\hat{A}}$ represents the estimated set of $\NFG$ smooth functions (often referred to as the mixing matrix) with shape [$N_\nu,\NFG$] that evolve the $\NFG$ separable source maps $\mathbf{S}$ through frequency. Generally the sources can be identified by projecting the mixing matrix along the observed data\footnote{For PCA and SVD, by construction, the set of functions identified for the mixing matrix are orthogonal and hence $(\mathbf{\hat{A}}^\text{T}\mathbf{\hat{A}})^{-1} = \mathbf{I}$; thus, this factor is often neglected in \autoref{eq:Sources} i.e. $\mathbf{S} = \mathbf{\hat{A}}^\text{T}  \mathbf{X}_\text{obs}$.}
\begin{equation}\label{eq:Sources}
    \mathbf{S} = (\mathbf{\hat{A}}^\text{T}\mathbf{\hat{A}})^{-1}\mathbf{\hat{A}}^\text{T}  \mathbf{X}_\text{obs}\,.
\end{equation}
$\NFG$ is the pre-selected number of separable sources which we expect our foreground emission to be contained within. The remaining signal from subtracting the smooth functions and sources from the data is in the residual term $\mathbf{R}$ and this is used for the cleaned intensity map data:
\begin{equation}\label{eq:HIlinearSys}
    \mathbf{X}_\text{clean} \equiv \mathbf{R} = \mathbf{X}_\text{obs} - \mathbf{\hat{A}}\mathbf{S}\,.
\end{equation}
This will contain cosmological \hinospace, noise and typically some residual foreground emission. The resulting cleaned intensity maps can be summarised as
\begin{equation}
	\delta T_\text{clean} = \delta T_\text{obs} - \widehat{\delta T}_\text{FG}(\nu,\boldsymbol{\theta}) = \delta T_\text{obs} - \sum_{n=1}^{\NFG} \hat{A}_{n}(\nu)\,S_n(\boldsymbol{\theta})\,.
\end{equation}
As we will see in this investigation, the optimal choice of $\NFG$ can vary considerably. Generally speaking, an $\NFG$ that is too low will result in too much foreground signal remaining in the residual component, and an $\NFG$ that is too high will result in too much cosmological \hi leakage into the subtracted component causing a loss of true signal. Finding an optimal balance is the aim of a successful foreground clean, and a key focus in our investigation. 

There are many existing methods for estimating $\mathbf{\hat{A}}$ for a given choice of $\NFG$, and we explore some in the remainder of this section. In this section we aim to introduce some of the most popular blind source separation techniques, and highlight their similarities. We will use an SVD-based technique (or equivalently PCA -- an equivalence we will explain) as our default foreground cleaning method which we introduce next. We then explore some related techniques with extended sophistication and test them on our simulated data.

We emphasise that the methods we outline are in no way an exhaustive list, and many more methods exist for foreground removal that could be applicable to 21cm intensity mapping e.g. GNILC \citep{olivari2016}, SMICA \citep{Delabrouille:2002kz}, RPCA \citep{Zuo:2018gzm} etc. (see the list in the Appendix of \citet{Leach:2008fi} for more information). One further notable approach is Gaussian Process Regression (GPR) \citep{Mertens:2017gxw}, which has been recently used on real data but for a higher redshift, epoch or reionisation survey \citep{Mertens:2020llj}. Investigating this method with low-redshift 21cm intensity mapping data is very interesting and will be the focus of future work. 

\subsection{PCA (\& SVD)}\label{sec:PCA}

Principal Component Analysis (PCA) is a widely used technique in statistics, closely related to Singular Value Decomposition (SVD). It provides a hierarchical coordinate system to represent high-dimensional correlated data by transforming it to a dimensional basis that maximises the variance. These new basis vectors are the principal components. In the context of correlated foreground emission in 21cm data, due to their large amplitude and highly correlated frequency structure, it is likely that the foreground signals can be reconstructed from just a few of these principal components. Hence, the first few $\NFG$ dominant basis vectors found in this process represent the estimate for the set of smooth functions in \autoref{eq:FGlinearSys}, which can then be removed from the observational data. 

The steps for performing PCA to construct an estimate of the foreground contamination $\mathbf{\widehat{X}}_\text{FG}$, which is then removed from the data, can be concisely outlined as follows:

\renewcommand{\theenumi}{\arabic{enumi}.}
\begin{enumerate}
    \item The data is mean-centred, i.e. the mean at each frequency is subtracted from the data for each frequency channel.\\
    \item The covariance matrix of the mean-centred data is calculated: $\mathbf{C} = \mathbf{X}_\text{obs}^\text{T}\mathbf{X}_\text{obs}/(N_\theta - 1)$.\\
    \item The eigen-decompositon of the covariance matrix is computed: $\mathbf{C}\mathbf{V}=\mathbf{V}\mathbf{\Lambda}$, where $\mathbf{\Lambda}$ is the diagonal matrix of eigenvalues ordered by descending magnitude.\\
    \item The first $\NFG$ column vectors $\boldsymbol{v}_i$ from the eigenvector matrix $\mathbf{V}$ represent the set of smooth functions to construct the [$N_\nu,\NFG$] mixing matrix i.e. $\mathbf{\hat{A}} = [\boldsymbol{v}_1,\boldsymbol{v}_2,\, ...,\,\boldsymbol{v}_{N_\text{FG}}]$.\\
    \item The projection of the selected eigenvectors along the mean-centred data provides the eigen-\textit{sources}, $\mathbf{S} = \mathbf{A}^\text{T} \mathbf{X}_\text{obs}$,
    which are combined with the mixing matrix to provide the reconstructed foreground estimation $\mathbf{\widehat{X}}_\text{FG} = \mathbf{A} \mathbf{S}$.
\end{enumerate}

\subsubsection{Singular Value Decomposition}\label{sec:SVD}

The singular value decomposition (SVD) is a unique matrix decomposition of the data (note that PCA and SVD are inherently related). The SVD of the observed data $\mathbf{X}$ is given by\footnote{The more general form for SVD is $\mathbf{X}=\mathbf{U} \mathbf{\Sigma} \mathbf{V}^\text{*}$, however in the context of 21cm data we are always dealing with real-valued matrices where $\mathbf{V}^\text{*} \equiv \mathbf{V}^\text{T}$.}
\begin{equation}\label{eq:SVD}
	\mathbf{X}=\mathbf{U} \mathbf{\Sigma} \mathbf{V}^\text{T} \, ,
\end{equation}
where $\mathbf{U}$ and $\mathbf{V}$ are unitary matrices with orthonormal columns and $\mathbf{\Sigma}$ is a diagonal matrix whose entries represent the \textit{singular values}. It can be demonstrated how closely related the SVD is to an eigenvalue decomposition. By considering \autoref{eq:SVD}, and given that $\mathbf{X}^\text{T} = \mathbf{V} \mathbf{\Sigma} \mathbf{U}^\text{T}$, the covariance can be written as
\begin{equation}
	\mathbf{C} \equiv \frac{\mathbf{X}^\text{T}\mathbf{X}}{N_\theta - 1} = \frac{\mathbf{V} \mathbf{\Sigma} \mathbf{U}^\text{T} \mathbf{U} \mathbf{\Sigma} \mathbf{V}^\text{T}}{N_\theta - 1} \, .
\end{equation}
A fundamental property of SVD stipulates that $\mathbf{U}^\text{T} \mathbf{U}$ is a unitary matrix ($\mathbf{U}^\text{T} \mathbf{U} = \mathbf{I}$). With a little rearranging we get
\begin{equation}
	\mathbf{C}\mathbf{V} = \frac{\mathbf{V} \mathbf{\Sigma}^2}{N_\theta - 1} \, ,
\end{equation}
which we can recognise as an eigenvalue decomposition of the correlation matrix $\mathbf{C}$ (as shown in step 3. in Section \secref{sec:PCA}) where the columns of $\mathbf{V}$ are the eigenvectors and the singular values $\mathbf{\Sigma}$ are proportional to the positive square roots of the eigenvalues.

In real intensity mapping data, in order to mitigate the high-levels of thermal noise \oldred{bias} present in pathfinder experiments \oldred{and decrease systematics}, it is necessary to cross-correlate data from different observation runs e.g. $\mathbf{X}_\text{A} \times \mathbf{X}_\text{B}$. \oldred{Whilst separating the data in this way decreases sensitivity and boosts thermal noise for each individual run, the noise should be uncorrelated for each run and thus thermal noise bias is mitigated in the cross-power.} In this situation, the covariance matrix $\mathbf{X}_\text{A}^\text{T}\mathbf{X}_\text{B}$ is no longer symmetric and an SVD is required where the left and right singular vectors in $\mathbf{U}$ and $\mathbf{V}$ are used to reconstruct foreground estimates in each run \citep{Switzer:2013ewa}. In this work we do not explore such a situation and therefore the SVD and PCA can be seen as equivalent treatments.

A related, and essentially equivalent, method to PCA is polynomial fitting \citep{Ansari:2011bv}. Although similarities exist, this is not to be confused with parametric fitting (see \secref{sec:NonBlindGMCA}) and conventionally refers to a blind approach to foreground cleaning. The approach works by identifying a set of smooth fitted functions $f_k$ where polynomials are used as basis functions e.g.
\begin{equation}
    f_k(\log (\nu))=[\log (\nu)]^{k-1}\,.
\end{equation}
Then, by least-squares fitting these functions to each line-of-sight, the foreground contribution can approximated. Since previous work has already demonstrated the theoretical equivalence this has with PCA (e.g. \citet{Alonso:2014dhk} that also provides simulation tests) we do not include this in our investigation.

\subsubsection{Truncation Choice}\label{sec:Truncation}

Deciding where to  truncate to, i.e. the number of $\NFG$ principal components to include in the foreground estimate (and hence remove) is key to an optimised blind foreground clean. By analysing the eigenvalues in $\mathbf{\Lambda}$ (or, equivalently, the singular values from the SVD), that estimate the amount of variance in the data captured in the corresponding principal components, an informed choice can be made. As discussed above, due to the nature of the foreground emission, most of the information is contained in a small sub-set of principal components where often $\NFG \sim 3 \rightarrow 20$ (depending on the foreground emission and instrument response) can produce a reasonable reconstruction. We can quantitatively analyse this choice with
\begin{equation}\label{eq:EigenvalRatio}
	R = \frac{\sum_{i=1}^{\NFG} \lambda_i}{\sum_{i=1}^{N_\nu} \lambda_i} \, ,
\end{equation}
where $\lambda_i$ are the eigenvalues in $\mathbf{\Lambda}$, descending in magnitude \red{and $N_\nu$ is the number of frequency channels along the line-of-sight}. Since a higher number for $\NFG$ will remove more \hi information, the aim for an optimal choice is to maximise $R\rightarrow 1$ for a minimal value for $\NFG$.

\begin{figure}
    \centering
    \includegraphics[width=\columnwidth]{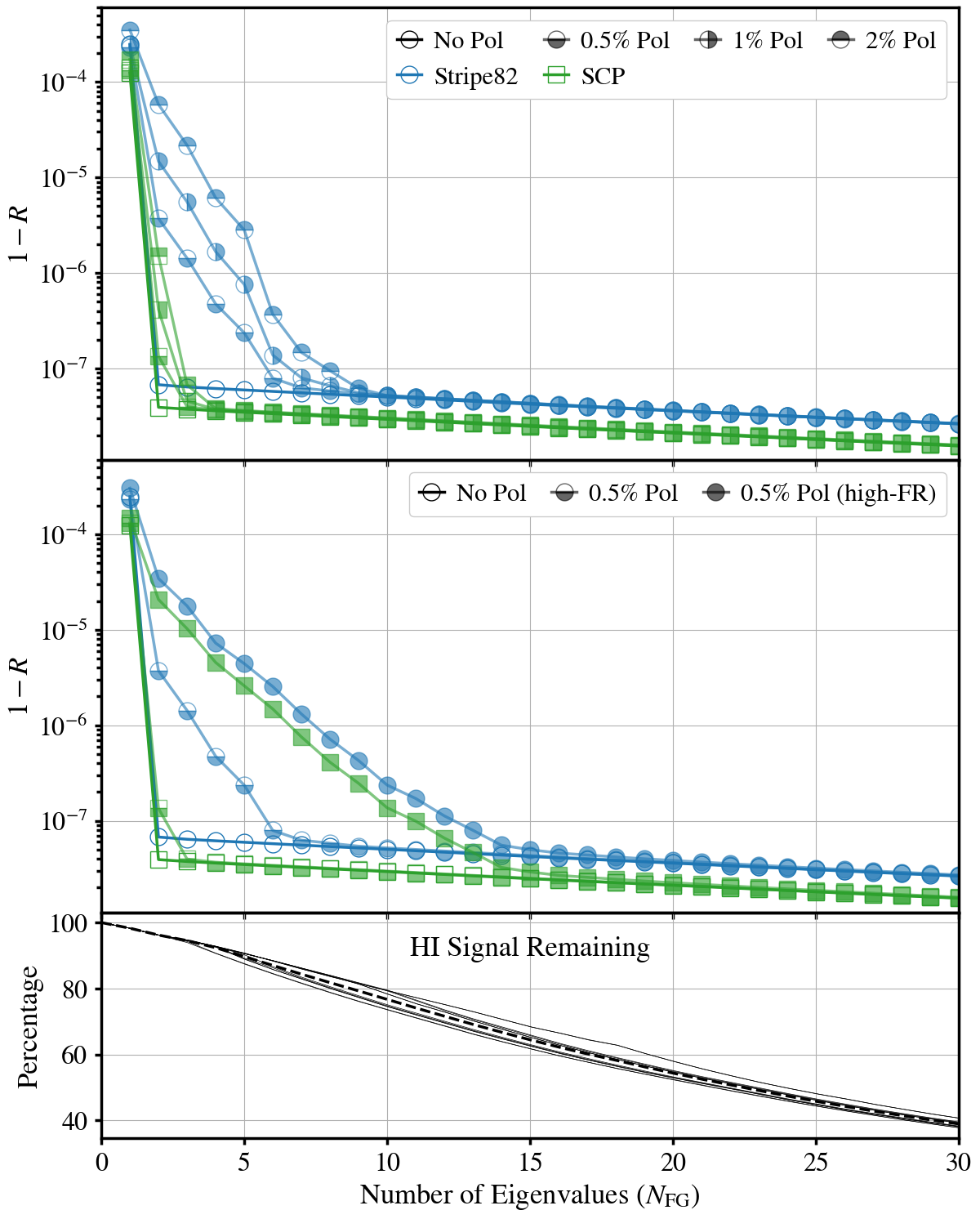}
    \caption{Weighted contributions from increasing numbers of principal components for the frequency-frequency covariance matrix for \oldred{different polarisation cases in both sky-regions.} $R$ (outlined by \autoref{eq:EigenvalRatio}) is the sum of the first $\NFG$ eigenvalues divided by the sum of all eigenvalues. Therefore the closer to zero $1-R$ is, the more eigen-information (or variance) is represented in those principal components. \oldred{(\textit{Top}-panel) the difference from an increasing polarised fraction percentage. (\textit{Middle}-panel) the impact from higher Faraday rotation (FR).} (\textit{Bottom}-panel) an estimation for the amount of \hi information along the line-of-sight that remains after $\NFG$ principal components are removed. Thin lines are for each of the \oldred{different cases in the above panels}, and the thick-dashed line is their average.}
    \label{fig:Eigenvals}
\end{figure}

We show some values for $R$ in \autoref{fig:Eigenvals} for the different sky regions in our simulated data, \oldred{with differing polarisation leakage cases}; we plot $1-R$ to demonstrate the convergence. This means the closer to zero the particular combination of eigenvalues is, the better a representation of the full data that reconstruction will be since it is capturing more of the full data's variance, and its reconstruction will capture more of the foreground contamination which can be removed. \autoref{fig:Eigenvals} immediately shows how highly correlated the observed data is given that just one eigenvalue in all cases has $R\sim 1$ meaning nearly 100\% of the signal can be represented with just 1 principal component ($\NFG=1$). However, just a small amount of residual foreground, even at the sub-percent level, is enough to entirely dominate the \hi signal. Therefore in all cases $\NFG=1$ is not sufficient for a foreground clean. This plot gives an indication of how far one needs to go in the reconstruction. All cases eventually reach a plateau where including more eigenvalues barely contributes to the reconstructed signal and it is here where PCA has likely reached its efficiency limit and will not be able to remove much more foreground. \red{The high Faraday rotation cases will demand the most modes for a successful reconstruction, requiring $N_\text{FG}\sim15$ to converge to the small eigenvalue plateau. We note that since this effect is being simulated from the Galactic plane for both Stripe82 and SCP regions, we will expect our results to converge to a plateau at a similar number of eigenvalues for both regions in this high-FR case. We see this in the middle panel and also see this in later results. This limits comparisons between the two regions for this extreme case but, as discussed, this approach is necessary to provide simulated data which require an aggressive foreground clean with a high $N_\text{FG}$.}

In contrast, the \hi information cannot be compressed into a small number of principal components due to its Gaussian-like nature. This is the main principle behind the blind source separation approach. The highly correlated information containing the majority of foregrounds can be removed using a few $\NFG$ modes, leaving the bulk of the \hi information that is mostly evenly distributed among the remaining components. However, it does mean a fine balance needs to be attained in a successful foreground clean. Being too aggressive and choosing too high values for $\NFG$ will begin to remove \hi information, typically large-scale line-of-sight modes. 

In  a simulation-based procedure, we can effectively analyse this problem since we have access to the separated \textit{pure}-\hi\footnote{We also include the contribution from thermal noise in this calculation.} and \textit{pure}-foreground simulated data. We can therefore calculate the contributions from these components remaining in the residuals after a foreground clean. The separated residuals are calculated using the estimated mixing matrix $\mathbf{\hat{A}}$, and projecting the pure-\hi (or pure-foreground) simulated data along this:
\begin{align}\label{eq:HIresids}
    &\mathbf{X}_\text{resid\hinospace} =   \mathbf{X}_\text{\hinospace} - \mathbf{\hat{A}}(\mathbf{\hat{A}}^\text{T}\mathbf{\hat{A}})^{-1}\mathbf{\hat{A}}^\text{T}  \mathbf{X}_\text{\hinospace}\,,\\
    \label{eq:FGresids}
    &\mathbf{X}_\text{residFG} = \mathbf{X}_\text{FG} - \mathbf{\hat{A}}(\mathbf{\hat{A}}^\text{T}\mathbf{\hat{A}})^{-1}\mathbf{\hat{A}}^\text{T}  \mathbf{X}_\text{FG}\,.
\end{align}
Note the $(\mathbf{\hat{A}}^\text{T}\mathbf{\hat{A}})^{-1}$ factor is not needed for the PCA method since, by construction, the mixing matrix is orthogonal and this quantity will equal the identity matrix. However, the vectors in the mixing matrix for the \fastica and GMCA approach we use are not orthogonal, thus this quantity is needed to obtain the correct projection.

We utilise these separated residual calculations extensively in our results to analyse the performance from various cleaning methods under different situations. We also use this concept for the bottom-panel of \autoref{fig:Eigenvals} where we demonstrate an estimation for the amount of eigen-information lost along the line-of-sight for each choice of $\NFG$. This calculated by computing the eigendecomposition of the residual \hi (\autoref{eq:HIresids}) and summing the eigenvalues. Dividing this by the sum of all the eigenvalues in the original \hi data gives a proxy for the amount of eigen-information remaining after subtracting $\NFG$ principal components, i.e. the portion of variance that is removed. This should further illustrate the challenge of foreground cleaning, a balance between removing foreground while trying to leave the \hi signal intact -- however, we always expect some signal loss. Since foreground dominated data is typically decomposed into dominant eigenmodes containing highly correlated information along the line-of-sight subtracting $\NFG$ principal components  generally removes large-scale modes along the line-of-sight in the \hi power spectrum, i.e. small $k_\parallel$ modes. Modelling this is non-trivial and a major challenge for precision radio cosmology. 

\subsection{\fastica}

Fast Independent Component Analysis (\fastica) is another widely used method for foreground cleaning and has been tested on simulated data \citep{Chapman:2012yj,Wolz:2013wna, Cunnington:2019lvb} and also on real data \citep{Wolz:2015lwa}. When we discuss \fastica\ we are referring to the method developed in \citet{Hyvrinen1999FastAR} and we use the package in \textit{Scikit-learn}\footnote{\href{https://scikit-learn.org/stable/modules/generated/sklearn.decomposition.FastICA.html?highlight=fastica}{https://scikit-learn.org/}} \citep{Pedregosa:2012toh}.

While PCA is generalised for reducing dimensionality in data, \fastica\ (and more generally independent component analysis) is more specifically used to separate mixed signals, and is therefore naturally suited to a blind source separation problem. \fastica\ forms an estimate for the mixing matrix $\mathbf{\hat{A}}$ by assuming the sources are statistically independent of each other. The method therefore aims to maximise statistical independence  that can be assessed using the central limit theorem, which states that the greater the number of independent variables in a distribution, the more Gaussian that distribution will be (that is, the probability density function of several independent variables is always more Gaussian than that of a single variable). Hence, by maximising any statistical quantity that measures non-Gaussianity, we can identify statistical independence.

Before assessing non-Gaussianity, \fastica\ begins by mean-centering the data then carries out a \textit{whitening} step that aims to achieve a covariance matrix equal to the identity matrix for this whitened data (i.e. the components will be uncorrelated and their variances normalised to unity). Since this whitening step can be achieved with a PCA analysis, \fastica\ is essentially an extension of PCA, and hence in most cases in the context of foreground cleaning, will provide very similar results.

For maximising non-Gaussianity, an approximation of the negentropy can be used\footnote{Kurtosis can also be used as a measure of non-Gaussianity.}. We refer the reader to \citet{HyvrinenOjaICA} for further detail on this aspect of the algorithm. In the context of 21cm foreground cleaning, the approximation of negentropy uses a set of optimally chosen non-quadratic functions which are applied to the data and averaged over for all available pixels. The maximisation of negentropy by averaging over angular pixels means that for purely Gaussian sources, \fastica\ will be unable to improve upon the initial PCA step carried out in the whitening step. This is because the Gaussian sources will have an equivalent zero negentropy. This explains the similarity in results often found between PCA and \fastica\ when most of the simulated components are Gaussian fields \citep{Alonso:2014dhk}. It is in situations over very large skies, where the negentropy approximation will be more optimal and sufficient non-Gaussian structure exists in the foreground maps, where \fastica\ will perhaps make discernible differences to the PCA-only performance.

To summarise, the components found using PCA are uncorrelated linear combinations of the data, which are identified by maximising the variance. \fastica\ extends on this by finding components that are also uncorrelated linear combinations of the data but identified by maximising statistical independence, through estimates of non-Gaussianity in angular pixels.

\subsection{GMCA}

GMCA stands for Generalised Morphological Component Analysis \citep{gmca}, it is a blind source separation algorithm exploiting the idea that the different components contributing to the signal are morphologically different. To enhance the morphological differences, the signal is projected into an adapted domain where we expect the components to have a sparse representation, i.e. to be described by few non-zero coefficients. When we find such a domain, the contrast between components increases, easing the separation process.
Here, we make use of wavelets, which has recently been shown to be optimal for this context \citep{Carucci:2020enz}. GMCA has already been optimised and used with astrophysical data sets (e.g. Cosmic Microwave Background data \citep{bobin2013,bobin2014}, high-redshift 21cm interferometric data \citep{Patil2017}, X-ray images of Supernova remnants \citep{Picquenot2019}).

In practice, once the data $\textbf{X}_\text{obs}$ has been wavelet-transformed to $\textbf{X}^{\rm wt}$, GMCA promotes sparsity in the requested $\NFG$ sources \textbf{S}$^{\rm wt}$ by solving iteratively the minimisation problem given by
\begin{equation} \label{eq:GMCAmaster}
    \{\mathbf{\hat{A}}, \mathbf{\hat{S}}\} = \min_{\textbf{A}, \textbf{S}^{\rm wt}}  \sum_{i=1} ^{\NFG}  \lambda_i \left\lvert \left\lvert \mathbf{S}^{\rm wt}_{i} \right\rvert \right\rvert_1+ \left\lvert \left\lvert \textbf{X}^{\rm wt} - \textbf{A} \mathbf{S}^{\rm wt}  \right\rvert \right\rvert_{F}^{2}, 
\end{equation}
where the first term is the $\ell_1$ norm, i.e. $\sum_{j,k} \left\lvert  \mathbf{S}^{\rm wt}_{j,k} \right\rvert$: this constitutes a constraint for sparsity, mediated by the regularization coefficients $\lambda_i$. The latter act as sparsity-thresholds that in our case should be tuned by the difference in intensity between the foregrounds and the cosmological signal; we first estimate them with the median absolute deviation (MAD) method and progressively decrease towards a final noise-related level. The second term in \autoref{eq:GMCAmaster} is the standard Frobenius norm, that assures data-fidelity step by step.

Once the mixing matrix $\mathbf{\hat{A}}$ has been estimated, we project the initial data $\textbf{X}_\text{obs}$ in pixel-space (following \autoref{eq:Sources} and \autoref{eq:HIlinearSys}) to retrieve the GMCA-reconstructed data cubes.  We refer the reader to \citet{Carucci:2020enz} for more details.

\subsection{Non-Blind Parametric Fitting}\label{sec:NonBlindGMCA}

In non-blind methods an estimator $\mathbf{\hat{A}}$ for the mixing matrix is constructed using astrophysical, as opposed to statistical, knowledge about the foreground sources and has been previously explored \citep{Ansari:2011bv,Bigot-Sazy:2015jaa}. A single frequency channel of a 21cm intensity mapping experiment will consist of diffuse synchrotron emission, diffuse free-free emission, extragalactic point sources, the \hi signal, instrumental noise and any other instrumental contributions (e.g. polarisation leakage). Synchrotron and free-free emission are believed to be spectrally smooth with well-understood spectral forms that can both be expressed as power laws. Whilst the synchrotron spectral index is known to change across pixels, diffuse synchrotron emission is the signal identified with the largest signal-to-noise ratio within 21\,cm intensity mapping experiments giving it the largest probability of an accurate characterisation. \oldred{It should be noted, however, that as synchrotron emission is around four orders of magnitude larger than the 21\,cm signal of interest it would need to be characterised to an accuracy of 0.01 per cent in order to no longer obscure the \hi signal.} 

\red{As such} we propose a parametric fit which aims to parameterize the free-free and synchrotron foreground contributions explicitly. Diffuse synchrotron, diffuse free-emission and extragalactic point sources are strongly degenerate; free-free and synchrotron emission maps contain identical spatial features and all three spectra can be represented as power laws with similar spectral indices. Whilst we do not aim to explicitly fit for the extragalactic point sources we expect their contributions to be subsumed within the synchrotron and free-free emissions fits. We are essentially making the opposite assumption to ICA by relying upon the parameter degeneracy, if this assumption is correct then the residuals between our parametric fit and the total data should contain the \hi plus any instrumental contributions. 

\citet{othparam} attempt a liner least-squares fitting to their data, modelling their combined foregrounds as a $n^{{\rm{th}}}$ order polynomial. We also use the least-squares optimiser (\autoref{eq:Sources}) for the emission sources. However, in an attempt to capitalise on existing foreground information, we aim to provide the optimisation with a realistic mixing matrix. We set up a mixing matrix with two components (to represent the combination of free-free, synchrotron emission and point sources). For the first component we use the assumption that the free-free spectral index is well-known and constant across pixels and so set the spectral form to the true value $(\nu/\nu_0)^{-2.13}$. For high Galactic latitudes such as the SCP and Stripe 82, the free-free emission is weak enough to be assumed negligible and thus we employ a least-squares fit assuming pure synchrotron emission. \oldred{Close to the Galactic Plane, we find that actual component separation between free-free and synchrotron emission is required. As intensity mapping experiments will not typically target such regions we shall not discuss parametric component separation any further. However, the interested reader can refer to \citet{camsap} for a description of a novel semi-supervised sparse component separation method, which has been used (by these authors and on the same total intensity simulations used in this work) to determine accurate synchrotron spectral indices in the presence of non-negligible free-free emission.}

Before describing our least-squares fit, we point out that the data monopoles must be removed from each map; for our particular simulations that means the unresolved extragalactic point source levels at each frequency. The spectral index for a particular emission is strongly tied to the monopole level of the maps and so the parametric fit we perform is tied to the zero-level of the observational data.

\oldred{For our least-squares fit, which assumes that synchrotron emission dominates the total intensity maps, we limit the parameter space for $\beta_\text{sy}$ to within $\pm 10\%$ of the total data spectral index and the parameter space for the synchrotron emission amplitude to within $\pm 50\%$ of the total temperature.}

\oldred{We also investigated the possibility of performing an MCMC fit (using \texttt{emcee} \citep{emcee}), to see if this offered significant benefits over least-squares fitting. We imposed flat priors on the synchrotron and free-free emission amplitudes and, following the methodology of \citet{erik}, the Jeffreys prior on the synchrotron spectral index. For the SCP region without polarisation leakage, a marginal (on average around a tenth of a per cent) improvement was seen for the estimation of the synchrotron spectral index. However, when polarisation leakage is added the MCMC fit no longer outperforms the least-squares fit. This is unsurprising as the strength of an MCMC Bayesian fitting process is the ability to provide end-to-end error propagation. In our case we are simply using the theoretical Gaussian noise level per frequency to form our noise estimates; we have no model for polarisation leakage as part of the noise estimates. As we aim to test how well a parametric fit of intensity mapping data can perform using existing astrophysical information we stick to least-squares fitting. Over time, however, it should be possible to perform an instrument specific, iterative Bayesian fit to intensity mapping data such as the CMB data analysis performed in \citet{beyondplanck}.}

Having used the least-squares fit to acquire the per-pixel synchrotron spectral index values we now have a complete mixing matrix which we use to calculate the diffuse Galactic emission amplitudes from the total temperature maps using \autoref{eq:Sources}. We can subtract our free-free emission and synchrotron emission estimates from the total temperature data to leave maps of HI plus instrumental contributions. 

\subsection{Quantifying Foreground Removal Effects}

Despite the range of different foreground cleaning methods available, none are perfect and will inevitably remove some cosmological \hi signal or leave behind foreground residuals. We discuss some methods for investigating this both with simulations and real data.

\subsubsection{Damping Cosmological \hi}\label{sec:TransferFunc}

This usually occurs on large scales where the \hi is most degenerate with the foregrounds. For idealised future surveys assuming excellent instrumental calibration, residual foregrounds should be well controlled and not exacerbated from effects such as polarisation leakage. In these cases the effects from a low-$\NFG$ foreground clean are relatively straightforward and can be potentially modelled as some damping to the power spectrum (see e.g. \citet{Cunnington:2020mnn,Cunnington:2020wdu,Soares:2020zaq}). 

However, applying this to real data requires a high level of confidence in the modelling that builds upon a detailed understanding of the nature of foregrounds as well as systematic/instrumental effects, something we do not currently have. An alternative approach is to add the observed data itself to simulations (mocks), then apply a foreground clean and access the response the mock data had to this process. Signal loss can be quantified this way with a \emph{foreground transfer function}, which is applied to the real data to compensate for these effects\oldred{, although does not avoid the reduced sensitivity caused by the contamination}. This has been the approach of several of the \hi intensity mapping detections so far (\citep{Masui:2012zc,Switzer:2013ewa,Anderson:2017ert,Wolz:2021ofa}).

\begin{figure}
    \centering
    \includegraphics[width=\columnwidth]{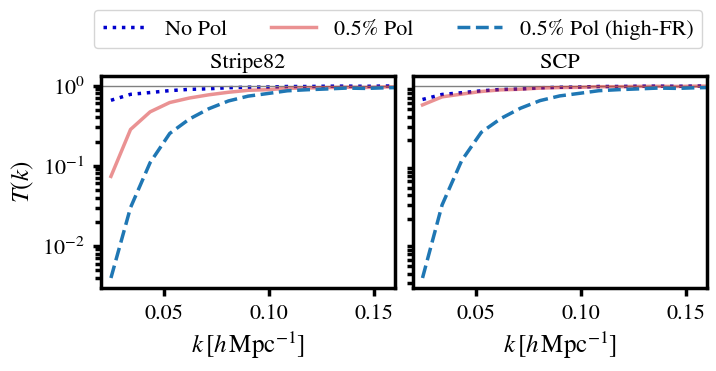}
    \caption{\red{Foreground transfer functions  for the different foreground cases in each sky region. These are produced using \autoref{eq:TransferFunc} and treating our \textsc{MultiDark} simulated intensity maps as "observed" data and then using 100 lognormal sims as the input mocks. For the three foreground cases (no polarisation leakage, 0.5\% polarisation leakage, and high Faraday rotation) we use a PCA clean with $N_\text{FG}=\{3,9,15\}$ for Stripe82 and $N_\text{FG}=\{3,4,15\}$ for SCP.}}
    \label{fig:TransferFunc}
\end{figure}

Following \citet{Switzer:2015ria} the transfer function can be constructed by adding mock data $\mathbf{M}$ to the true observed data $\mathbf{X}_\text{obs}$, which includes foregrounds. This can then be cleaned to provide $\mathbf{M}_\text{cleaned}$, an estimate for the effects of removing the foregrounds on the mock map:
\begin{equation}\label{eq:MockforTF}
	\mathbf{M}_\text{cleaned} = [\mathbf{M} + \mathbf{X}_\text{obs}]_\text{PCA} - [\mathbf{X}_\text{obs}]_\text{PCA}\, .
\end{equation}
where the $[\ ]_\text{PCA}$ notation represents performing a PCA clean, but in principle this could be done with any foreground cleaning method. Note that in \autoref{eq:MockforTF} the cleaned data $[\mathbf{X}_\text{obs}]_\text{PCA}$ has been subtracted. This is necessary to reduce the variance in this estimation since the unwanted data-\hi component will serve as additional, unwanted noise. The transfer function is then given by:
\begin{equation}\label{eq:TransferFunc}
	T(k) = \left\langle  \frac{\mathcal{P}(\mathbf{M}_\text{cleaned}\, ,\, \mathbf{M})}{\mathcal{P}(\mathbf{M}\, ,\, \mathbf{M})} \right\rangle^2 \, ,
\end{equation}
where $\mathcal{P}()$ denotes an operator which measures the power spectrum in ($k_\perp,k_\parallel$) space. The angled brackets denote an averaging over a large number of mocks. The power spectrum is then corrected for by dividing through by this transfer function. This can also be utilised in a cross-correlation measurement with the only difference being that the power of 2 is dropped from \autoref{eq:TransferFunc} because the effects of cleaning are only applied to the \hi data. We employ the transfer function later in our analysis (\secref{sec:CrossCorrResults}) by constructing 100 lognormal mocks and using our \textsc{MultiDark} simulations as the "observed" data. \red{We show the transfer functions for the different data sets in \autoref{fig:TransferFunc}. This shows the range in foreground contamination from the different foreground cases, largely driven by the choice of $N_\text{FG}$ components to remove, which we discuss in detail in our results (\secref{sec:Results}).}

The foreground transfer function is thus used as a data-driven way of compensating for signal loss in the foreground removal. \oldred{Since the \textit{real} data is used in its construction, it should incorporate the interplay between foreground and systematics (even unknown systematics). Some assumptions are inherently made regarding the degeneracy of the transfer function with cosmological parameters, but this will only be a problem for precision cosmology which will need more detailed investigation when \hi intensity mapping reaches this level. This subtlety regarding degeneracies is discussed further in \citet{Cunnington:2020mnn}.} Since $T(k)\leq1$, the transfer function is not capable of addressing the issue caused by \emph{additive} biases from foreground residuals, discussed in the following section.

\subsubsection{Foreground Residuals}\label{sec:FGresid}

Whilst signal loss from over-cleaning can be modelled or compensated for with a foreground transfer function, foreground residuals produced from under-cleaning, which cause additive biases and boost errors, are more challenging to address. For near-future, pathfinder surveys (e.g. MeerKAT \citep{Santos:2017qgq}) it is possible that the instrument response will not be sufficiently understood and polarisation leakage effects could manifest, causing contamination from foreground residuals. Developing robust statistics which estimate the effects caused by these residuals will therefore be essential for future surveys \citep{Switzer:2015ria}. There is not a large amount of research on this issue, since current data analysis usually has large thermal noise and unknown systematic effects \citep{Switzer:2013ewa}. Alternatively, detections have been made using cross-correlations with optical galaxy data (e.g. \citet{Masui:2012zc,Anderson:2017ert,Wolz:2021ofa}). In cross-correlation the residual foregrounds and survey-specific systematics do not correlate with the optical galaxy data and instead, simply boost errors (we will study this in detail in \secref{sec:CrossCorrResults}). As the intensity mapping technique matures and calibration and signal-to-noise capabilities of surveys improve, we will aim to conduct precision cosmology using auto-correlation measurements. Therefore we need to develop a pipeline for quantifying the foreground residual contamination. 

As discussed in \secref{sec:Truncation}, the residuals can be exactly calculated (see \autoref{eq:HIresids} and \autoref{eq:FGresids}) because we are using simulations where the original decomposed \hi and foreground contributions are known. A direct comparison between $\mathbf{X}_\text{resid\hinospace}$ and $\mathbf{X}_\text{residFG}$ is then extremely useful (and a topic we investigate) where one ideally desires a situation where $\mathbf{X}_\text{resid\hinospace}$ dominates $\mathbf{X}_\text{residFG}$. A more dominant $\mathbf{X}_\text{residFG}$ would increase additive biases due to the residuals correlating with each other. However, in real data, distinguishing the contribution between foreground residual and \hi signal will be challenging. If one can develop a robust way of estimating the contribution from foreground residuals, then this can be effectively modelled in a similar way to the instrumental noise which causes additive power in auto-correlations along with boosting errors. So the \hi auto-power spectrum could be expressed as
\begin{equation}\label{eq:PowerWithResidFG}
    P_\hinospace(k) = \langle T_\hinospace \rangle^2 b^2_\hinospace P_\text{m}(k) + P_\text{N}(k) + P_\text{residFG}(k)\,,
\end{equation}
where $P_\text{m}$ is the matter power spectrum and $P_\text{N}$ and $P_\text{residFG}$ are the contributions from thermal noise and residual foregrounds. An estimation for the errors can then be analytically made with
\begin{equation}\label{eq:PowerErrorResidFG}
    \sigma_P(k) \sim \frac{P_{\hinospace}(k) + P_{\text{N}}(k) + P_{\text{residFG}}(k)}{\sqrt{N_\text{modes}(k)}}\,,
\end{equation}
where $N_\text{modes}$ is the number of unique modes in each $k$-bin, included to account for cosmic variance.

\section{Results}\label{sec:Results}

\begin{figure}
    \centering
    \includegraphics[width=\columnwidth]{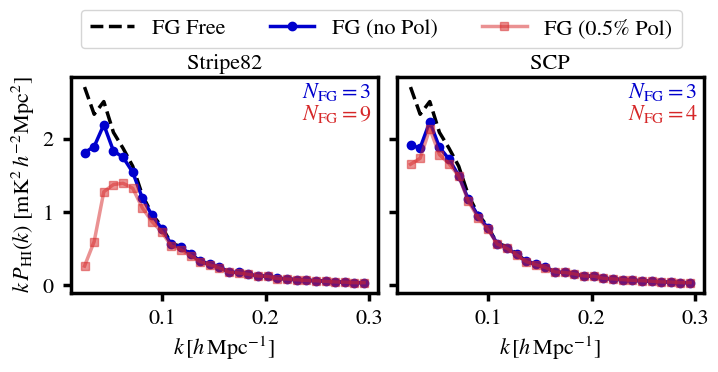}
    \caption{Measured power spectra for \oldred{both regions}. We show the foreground free case (\textit{black-dashed}), cleaned foregrounds without polarisation leakage (\textit{blue-circle} markers) and cleaned foregrounds with \oldred{0.5\%} polarisation leakage (\textit{red-square} markers). \oldred{For both regions where there is no polarisation leakage, we use $\NFG=3$ as the number of removed principal components, but for the polarisation leakage cases, a differing selection is needed} (displayed in top-right of both panels). In all foreground cleaned cases, PCA is used.}
    \label{fig:PkPatch}
\end{figure}

Here we present our results from tests carried out on the simulated data sets and foreground removal methods outlined in the previous sections. To diagnose the performance of our foreground cleans we look at measurements of power spectra, both 1D $P(k)$ and 2D $P(k_\perp,k_\parallel)$, and compare these to equivalent foreground-free results where only the cosmological \hi is being measured. The offset between the two then serves as a good indicator for how well the chosen method is performing. Following previous studies (e.g. \citet{Alonso:2014dhk,Carucci:2020enz}), we define below the weighted difference between subtracted foreground and no foreground cases as a metric to help assess the success of the foreground removal under all the different scenarios:
\begin{equation}\label{eq:PkEps}
    \varepsilon(k) = \frac{P_\text{SubFG}(k) - P_\text{NoFG}(k)}{P_\text{NoFG}(k)}\, .
\end{equation}
Here $P_\text{SubFG}(k)$ is the measured power spectrum for the simulated intensity maps with foregrounds included and then cleaned, while $P_\text{NoFG}(k)$ is the measured power spectrum of the \hinospace-only (foreground-free) intensity maps. We also analyse the 2D power spectrum and use an identical analysis in this basis where
\begin{equation}\label{eq:2DPkEps}
    \varepsilon_\text{2D}(k_\perp,k_\parallel) = \frac{P_\text{SubFG}(k_\perp,k_\parallel) - P_\text{NoFG}(k_\perp,k_\parallel)}{P_\text{NoFG}(k_\perp,k_\parallel)}\, .
\end{equation}
We begin by plotting the auto-power spectra for \oldred{both our chosen regions} and perform a comparison between the foreground-free \hi power spectrum (black dashed line) and the foreground-cleaned results using the PCA method. This demonstrates some differences between the regions and also some clear differences for the cases with polarisation leakage (red squares) and without (blue circles). As expected most of the damping to the power comes from large scales (small-$k$). Because we use $\NFG=3$ in \oldred{both regions where there is no polarisation leakage}, the damping across all regions is approximately equal. However, the foreground residuals could still differ in each region, with the most likely case being that residuals will be highest in \oldred{Stripe82 where foregrounds are slightly more dominant}. \autoref{fig:PkPatch} lastly shows that results can be extensively worse when including polarisation leakage effects, except perhaps the SCP, the region least affected by polarisation \oldred{in the \texttt{CRIME} model. Results are much worse for Stripe82, where we opted to use $\NFG=9$ to remove more oscillating foreground contamination from polarisation leakage. For example the largest mode (smallest-$k$) for Stripe82 is effectively damped to zero. We discuss in more detail the choice of $\NFG$ later in \secref{sec:OptimalNFG}.}

\begin{figure*}
    \includegraphics[width=2.1\columnwidth]{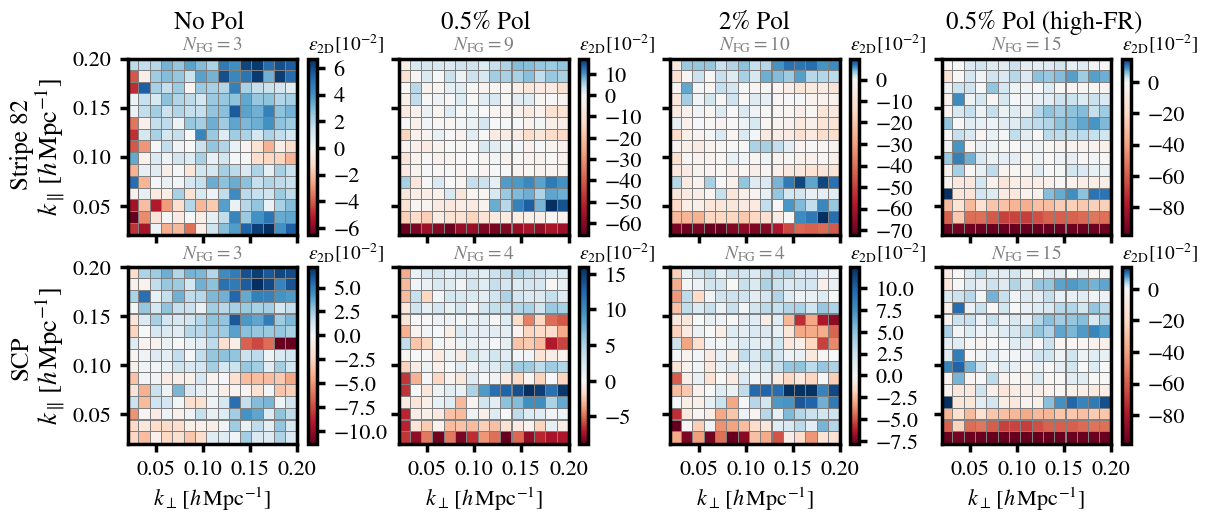}
    \caption{The impact of a PCA foreground clean on the 2D power spectra. Plotted is the weighted difference $\varepsilon_\text{2D}(k_\perp,k_\parallel)$ between the foreground free and cleaned 2D power spectra (outlined in \autoref{eq:2DPkEps}). Each panel represents a different \oldred{polarisation leakage case for both Stripe82 (\textit{top}-row) and SCP (\textit{bottom}-row) sky regions. We show 0\%, 0.5\% and 2\% leakage along with the special high-Faraday rotation (FR) case.} The results can be loosely interpreted as blue positive-pixels representing \textit{under}-cleaned modes (modes with dominant foreground residual), and red negative-pixels representing \textit{over}-cleaned modes. The values for $\NFG$ are \oldred{displayed in grey above each panel.}}
    \label{fig:2DPkPatch}
\end{figure*}

We show similar results in \autoref{fig:2DPkPatch} but here more information can be extracted on the nature of this contamination. This shows the weighted difference $\varepsilon_\text{2D}$ between the \hinospace-only 2D power spectrum and the foreground cleaned one. Again we plot \oldred{both regions on different rows but now with a wider range of polarisation leakage cases}. This gives an illustration into the effects of foreground under-cleaning and over-cleaning and the delicate balance between the two. From \autoref{eq:2DPkEps} we can see that the blue regions are indicating modes that have higher power in the foreground-cleaned maps compared to the foreground-free ones. Thus blue areas indicate under-cleaned $k$-space. Conversely, red areas have lower power in the foreground-cleaned maps, indicating over-cleaning. 

\autoref{fig:2DPkPatch} therefore shows the effects of over-cleaning tend to manifest in low-$k$ modes, as expected. This is particularly evident when going to high $\NFG$ as is \oldred{required in the high Faraday rotation polarisation leakage cases (far-right column)}, where we see significant damping to low $k_\parallel$ modes, again as expected. This is because, in order to control contamination from polarisation leakage, we are removing more principal components, each with different oscillating modes due to the instrumental response, but all will still have largely frequency correlated spectra. This inevitably removes the modes in \hi which are also highly correlated in frequency i.e. low-$k_\parallel$ modes. This conclusion is quite general and not just specific to a PCA-based method. Generally, any method that utilises the highly correlated nature of foreground signals will struggle to disentangle foregrounds and large \hi modes parallel to the line-of-sight.

For the unpolarised cases, where a lower $\NFG$ is used, it it interesting to see that small-$k_\perp$ modes are damped to a similar level as the small-$k_\parallel$ modes. This will be because the foregrounds generally have quite large angular structures (see maps in \autoref{fig:regions}), therefore when removing the dominant principal components representing the foregrounds, any small-$k_\perp$ modes will be degenerate with these and are therefore damped. This agrees with results from \citet{Soares:2020zaq} where a damping to $k_\perp$ modes was required to model the foreground contamination. 

\oldred{Comparing the middle-left and middle-right panels in \autoref{fig:2DPkPatch}, the increased polarisation fraction does not appear to have a drastic effect. This is also consistent with the results from the top-panel of \autoref{fig:Eigenvals} which showed that an increase in polarisation fraction, seems to just increase the amplitude of the eigenvalues one would remove anyway in a foreground clean. A polarisation fraction above 2\% is unlikely to occur in surveys, even in the most pessimistic of forecasts and we therefore stick to using a polarised fraction of 0.5\% for the remainder of the paper. However, as we demonstrate in the far-right panels, one way to complicate foreground cleaning is if Faraday rotation is higher than is being realised in the \texttt{CRIME} model. As explained in \secref{sec:PolarisatitonLeak}, the way we investigate the possibility of a more complex frequency decoherence situation is by taking the \texttt{CRIME} polarisation output map from the Galactic plane, and using this as a high Faraday rotation case. Doing this creates much more oscillating behaviour in the foreground spectra and as \autoref{fig:2DPkPatch} shows, requires a more aggressive clean, with $\NFG=15$.}

\subsection{Comparing Blind Foreground Cleaning Methods}\label{sec:BlindComparison}

\begin{figure}
    \centering
    \includegraphics[width=\columnwidth]{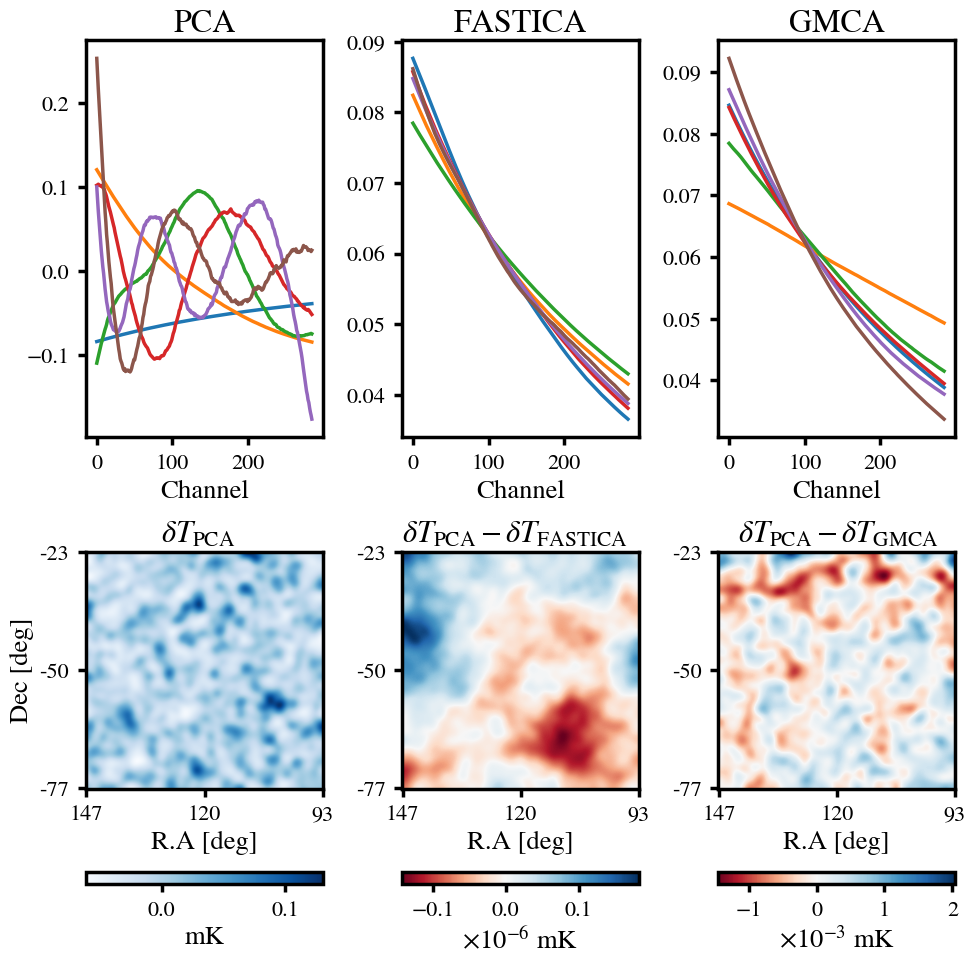}
    \caption{\textit{Top panels:} Normalised column vectors of the mixing matrix estimated by PCA, \fastica\ and GMCA for the \oldred{Stripe82} region with \oldred{0.5\%} polarisation leakage. Note that we only show the first 6 modes for clarity despite using \oldred{$\NFG=9$}. \textit{Bottom-left}: resulting \hi intensity map from PCA clean for a random frequency channel. Also shown are the difference maps between this PCA cleaned map and those cleaned using \fastica\ (\textit{bottom-centre}) and GMCA (\textit{bottom-right}). The three methods estimate different mixing matrices, yet they lead to analogous foreground-dominated data cubes and produce extremely similar \hi residuals as shown by the bottom maps \oldred{(noting the small colour-bar scale)}.}
    \label{fig:mixmat}
\end{figure}

We now compare the cleaning methods we have introduced: PCA, \fastica\, and GMCA. All methods rely on the assumption that we can decompose the signal linearly as in \autoref{eq:FGlinearSys} and estimate the mixing matrix $\mathbf{\hat{A}}$, identifying the subspace of the data set where we expect foregrounds to live, which are then removed as per \autoref{eq:HIlinearSys}.

\begin{figure}
    \centering
    \includegraphics[width=\columnwidth]{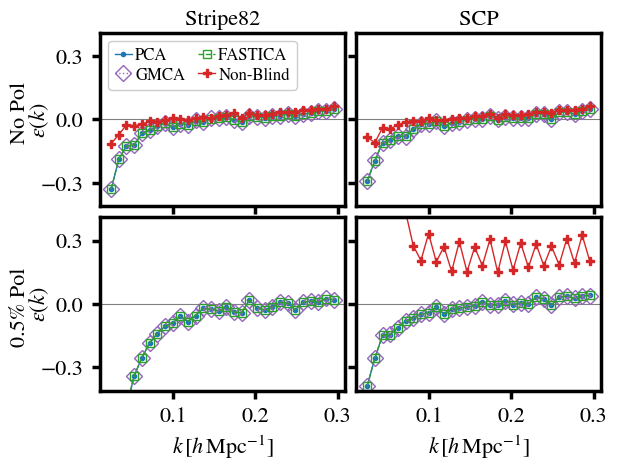}
    \caption{Impact from different foreground cleaning methods on the 3D power spectra \oldred{for both regions}. We show the weighted difference $\varepsilon$ in foreground free and cleaned power spectra (\autoref{eq:PkEps}) where positive (negative) values represent under- (over-)cleaning. Results are \textit{without} polarisation leakage (\textit{top-row}) and with 0.5\% polarisation leakage (\textit{bottom-row}).}
    \label{fig:FGremovalcomparison}
\end{figure}

In the top panels of \autoref{fig:mixmat} we show the mixing matrices derived by the different methods applied to the same data cube. Each method provides a different estimation for $\mathbf{\hat{A}}$, yet, the final cleaned maps are remarkably similar in all cases. We found that there were no discernible differences in the maps and in the bottom panel we demonstrate this. We plot the PCA-cleaned \hi intensity map for one channel (bottom-left panel of \autoref{fig:mixmat}) and then show the difference between this PCA map and the corresponding \fastica\ and GMCA counterparts (middle and right bottom panels). Their differences are orders of magnitude below the amplitudes of the cleaned map and this is true for all channels of all the sky regions explored, with and without the inclusion of polarisation leakage. Given the similarity in the cleaned maps, it is unsurprising that at the 2-point statistics level, in all the scenarios considered, the three blind methods output essentially identical power spectra (as shown by the examples in \autoref{fig:FGremovalcomparison}). 

The difference in the top-panel between PCA and the other methods is an interesting demonstration of the subtle distinction in techniques. PCA is maximising the variance into as few modes as possible. The highest ranked mode represents the one that best fits the variance of the data. The second highest rank mode, will be the next best fit but is required to be orthogonal to the first, hence why PCA identifies two dominant smooth modes in \autoref{fig:mixmat}, which likely contain the synchrotron and free-free emission. The remaining modes are then more oscillatory and likely identify polarised residuals in a descending order of contribution to the total variance. Applying \fastica\ and GMCA algorithms, we are instead identifying a pre-determined $\NFG$ number of modes within which to maximise statistical independence or sparsity. They achieve this by identifying functions that can share out the contributions to the variance amongst these $\NFG$ modes, with no requirement of orthogonality, providing they are maximising independence and sparsity.
This is allows the modes in \fastica\ and GMCA to approximately follow the slope defined by the dominant spectral indices from synchrotron and free-free emission. These functions will still contain the polarised information demonstrated in the PCA functions, but be contained as sub-dominant oscillations within these modes.

Despite the differences in the identified mixing matrices, the similarity in final results from all three blind methods can be understood by considering the difference in the assumptions they make when linearly-decomposing the signal. PCA merely identify the eigenvectors corresponding to the largest eigenvalues in the frequency-frequency covariance of the data. \fastica\ adds on this by promoting non-Gaussianity in the estimated sources, as a proxy for their statistical independence. GMCA promotes sparsity in the spatial domain, after having wavelet-transformed the patches, relying on highlighting the specific morphologies of the components to facilitate the source separation process. Since all three approaches result in essentially identically cleaned maps and power spectra, this leads to the conclusion that the \fastica\ and GMCA assumptions do not hold for these data sets: we are dealing with fairly Gaussian (non-sparse) components in the spatial dimensions, thus \fastica\ and GMCA are not optimised to improve upon their pre-processing PCA step. However, this statement cannot be generalised and is specific to our tested simulated data i.e., the size and resolution of the patches we work with. For instance, \citet{Carucci:2020enz} show how \fastica\ and GMCA behave differently in presence of non-continuous, RFI-flagged data on the full-sky. Also work is still needed to understand a realistic beam effect on the intensity maps, which could add complexity to the foreground removal process.

\subsection{Non-Blind Foreground Cleaning}\label{sec:NonBlindResults}

\begin{figure*}
    \centering
    \includegraphics[width=2.1\columnwidth]{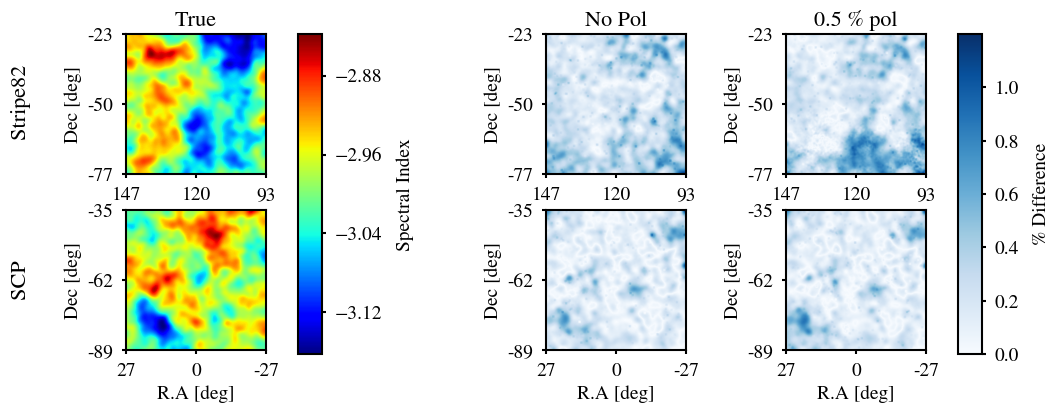}
    \caption{\oldred{True synchrotron spectral indices (\textit{left}-panel) from our simulations and the absolute percentage difference between the truth and those estimated by parametric fitting for no polarisation leakage (\textit{centre}-panel) and 0.5\% polarisation leakage (\textit{right}-panel). The Stripe82 region is on the \textit{top}-row with SCP on the \textit{bottom}-row.}}
    \label{fig:nonblindspecs}
\end{figure*}

The appeal of a parametric method is its ability to yield estimates for both the cosmological signal and each individual foreground. In our parametric approach we have assumed knowledge of the free-free spectral index, enlisted \oldred{a least-squares fit assuming pure synchrotron emission} to determine the synchrotron spectral index, and finally solved a least-squares optimization to the data based on the synchrotron and free-free spectral forms. As our approach only considers synchrotron and free-free emission explicitly, all other data contributions get absorbed into either our cosmological, synchrotron or free-free estimates. Specifically, the instrumental noise is included in our \hi estimate, point sources are contained in our synchrotron estimate due to their similar spectral forms and polarisation leakage is seen to degrade the quality of the \hinospace, free-free and synchrotron estimates.

With the free-free spectral form held at the true value, the key to our approach is accurate determination of the synchrotron spectral indices. \autoref{fig:nonblindspecs} shows the absolute percentage difference maps between the true synchrotron spectral indices and those estimated by our non-blind fit. \oldred{The right-hand column shows a slight increase in percentage error when polarisation leakage is present. However, it appears it is only a mild prohibitive factor for accurate recovery of the synchrotron spectral index in the high Galactic latitude regions we have studied and for the polarisation leakage model we have used.}

\oldred{We find parametric fitting is competitive with the blind methods in the absence of polarisation leakage. However, as polarisation leakage is not explicitly accounted for within our parametric fit, nor does it display any degenerate behaviour with the foregrounds we do account for, we find that parametric fitting is not capable of being used "as is" to get to the \hi signal level. This is explicitly shown in \autoref{fig:FGremovalcomparison}; in the case of the polarised Stripe82, the non-blind residuals are too large to occupy the same axis ranges as the blind residuals. In conclusion: our parametric fit cannot compete with the blind methods in the face of polarisation leakage.} 

\subsection{Hybrid Foreground Cleaning}\label{sec:Hybrid}

\begin{figure}
    \centering
    \includegraphics[width=\columnwidth]{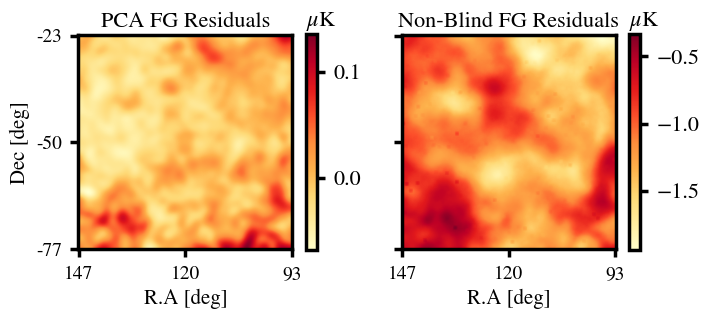}
    \caption{\oldred{Pure foreground residuals remaining after PCA (\textit{left}-panel) and parametric (\textit{right}-panel) cleans on Stripe82 region without polarisation leakage. We have averaged along the lines-of-sight.}}
    \label{fig:FGresidMaps}
    \centering
    \includegraphics[width=\columnwidth]{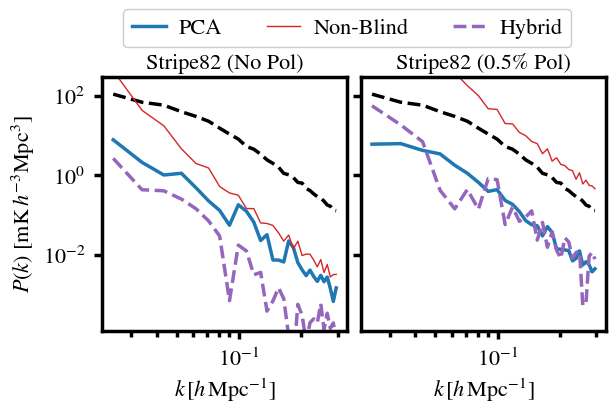}
    \caption{Power spectra for foreground residuals remaining after cleaning relative to the foreground-free original \hi signal \textit{(black-dashed line)}. Hybrid result refers to the cross-power spectrum between a PCA clean map and one cleaned using our non-blind approach.}
    \label{fig:HybridFGresids}
\end{figure}

With several available methods for performing 21cm foreground cleaning, an obvious question to ask is whether any of them can be combined into a hybrid method to produce better results. Hybridised techniques have been explored in \citet{Akrami:2018mcd}, where SMICA was used in a semi-blind way \oldred{and in \citet{camsap}, where GMCA was used in a semi-blind way}

The approach we adopt here differs slightly since we investigate combinations of foreground cleaning methods by cross-correlating two maps cleaned using different approaches. The main potential benefits from this approach come from the possibility of the cross-correlation reducing the residual foregrounds.

Due to the inherently similar resulting maps in our three blind methods, it is not surprising that  we found no benefit in combining these techniques. However, for cases where the non-blind approach was providing a good fit to  the foregrounds and thus producing a reliable foreground clean, we found that a cross-correlation between a map cleaned with this approach and one using PCA, does produce a reduction in residual foreground correlation. \oldred{We plot the pure foreground residuals (\autoref{eq:FGresids}) from PCA and the non-blind method for the Stripe82 region with no polarisation leakage in \autoref{fig:FGresidMaps}. It is clear by eye that there are some significant differences in these maps and thus cross-correlating these methods should result in a reduction from residual foreground contributions. We demonstrate this in \autoref{fig:HybridFGresids} (left-panel) showing the contribution to the power from the pure-foreground residuals} left after a clean using PCA (thick blue line) and parametric fit (thin red line). The aim is for the foreground residuals to be as low as possible, ideally far below the original \hi signal (black dashed line), which we include for reference. As shown, by using a hybrid approach (dashed purple line) that cross-correlates the two differently cleaned  maps, foreground residuals are reduced, \oldred{although we find no such reduction in the polarised case (right-panel)}. We found no discernible improvement in the final power spectrum measurement of all the components, i.e. this current approach would make no change to the measured $\varepsilon$. However, this reduction in residuals is important, especially for future surveys where maximising precision and reducing error-bars is paramount (see discussion in \secref{sec:FGresid}). This technique could also yield benefits in more realistic situations containing more systematics. If the different methods respond differently to these systematics, then reductions in residual contamination could be significant. For now we highlight these results as an area for potential further investigation.

\subsection{Balancing Foreground Residuals \& \hi Damping}\label{sec:OptimalNFG}

\begin{figure*}
    \centering
    \includegraphics[width=2.1\columnwidth]{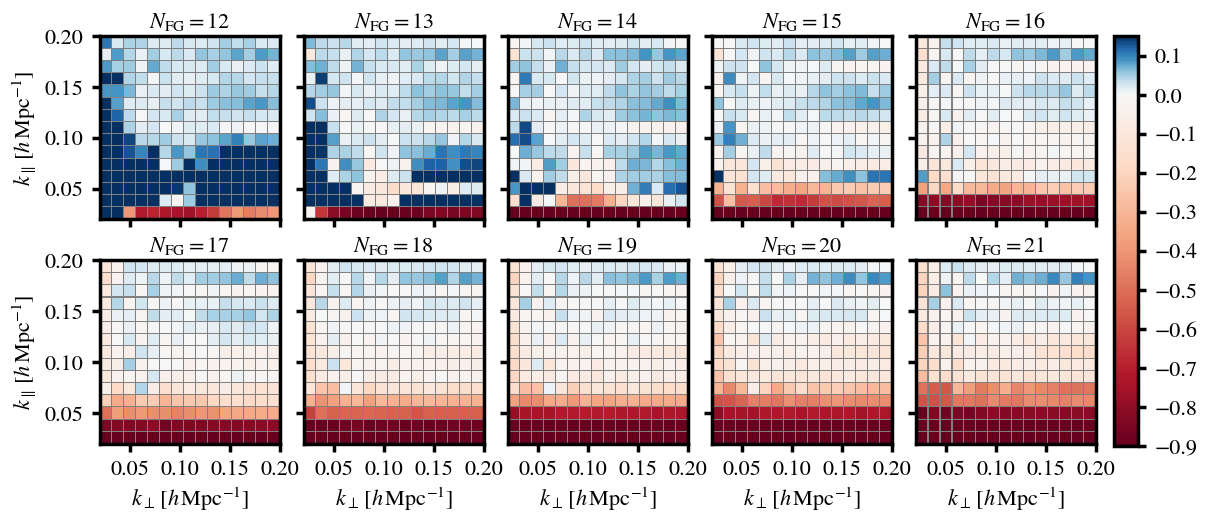}
    \caption{Impact of a varying $\NFG$ on the efficacy of the foreground clean. We show the weighted difference $\varepsilon_\text{2D}$ in foreground free and cleaned 2D power spectra, defined by \autoref{eq:2DPkEps}, where positive (negative) values represent under- (over-)cleaning. This is shown for a range of $\NFG$ values in a PCA clean on the \oldred{Stripe82 region for the high-Faraday rotation polarisation leakage case.} We see too low $\NFG$ leaves large foreground residuals and increasing $\NFG$ removes these residuals at the cost of damping small $k_\parallel$ modes.}
    \label{fig:NFGcomparison}
\end{figure*}

As intensity mapping surveys continue to produce data, the focus will be on how to optimise these surveys for constraining cosmological and astrophysical parameters. Until now we have used a consistent defined choice of $\NFG$ for each region and polarisation case. Here we begin to examine the consequences of varying this parameter and explore whether an optimal choice can be made. We seek a balance between over-cleaning foregrounds by using a high-$\NFG$ that causes damping to \hi power on large scales, and under-cleaning using a low-$\NFG$ leaving higher residual foregrounds that potentially bias results or boost errors.

In the case of cross-correlations (which we focus on in \secref{sec:CrossCorrResults}) this is more straightforward to analyse since residual foregrounds, provided they are not too large, will just boost the errors in the power spectrum measurement. Thus identifying an optimal number of $\NFG$ modes to remove is primarily based on minimising errors. In auto-correlation, the process for optimizing the choice of $\NFG$ is more difficult (see discussion in \secref{sec:FGresid}).

\autoref{fig:NFGcomparison} shows how the 2D power spectrum evolves with an increasing $\NFG$ by analysing $\varepsilon_\text{2D}$, the weighted difference between foreground-cleaned and foreground-free power spectra (\autoref{eq:2DPkEps}). The results are only for the \oldred{Stripe82 region in the case of high Faraday rotation (FR) polarisation leakage}, cleaned using a PCA method. This demonstrates how increasing the aggressiveness of the clean mitigates foreground residuals (blue regions) but at the expense of severely damping small-$k_\parallel$ modes (shown by red regions).

\begin{figure*}
    \centering
    \includegraphics[width=2.1\columnwidth]{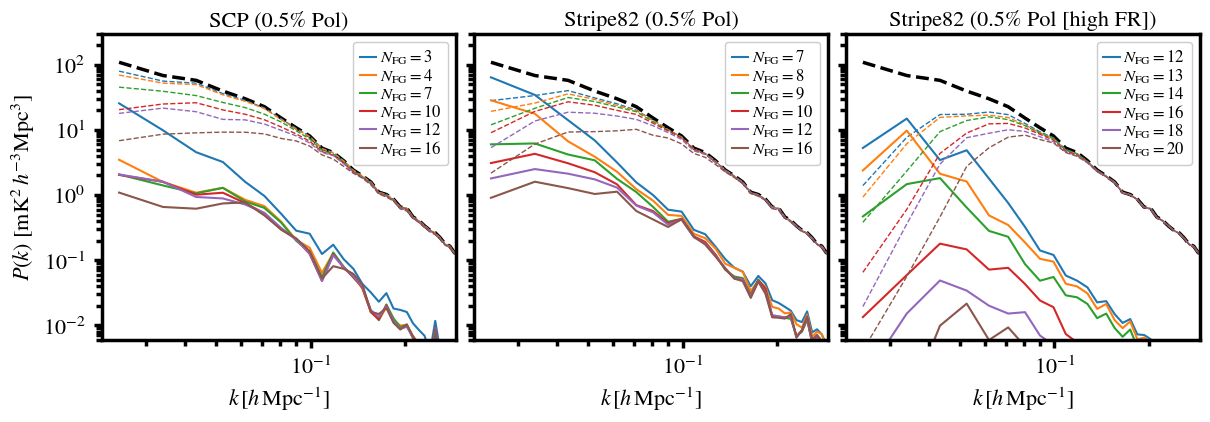}
    \caption{Power spectra for the foreground residuals (\textit{solid} lines) from a PCA clean with varying $\NFG$ modes removed from each region. \oldred{\textit{Left} and \textit{centre} panels are for both regions with 0.5\% polarisation leakage and the \textit{right} panel is the Stripe82 region for the high Faraday rotation case.} For comparison, we also plot the foreground-free original \hinospace-only signal (\textit{black-dashed} line) along with the residual \hinospace-only power spectra (\textit{coloured-dashed} lines) after each PCA clean. Ideally one would require a scenario where the residual \hi dominates over the residual foregrounds. Calculations outlined in \autoref{eq:FGresids} and \autoref{eq:HIresids}.}
    \label{fig:ResidualFGs}
\end{figure*}

\begin{figure*}
    \centering
    \includegraphics[width=2.1\columnwidth]{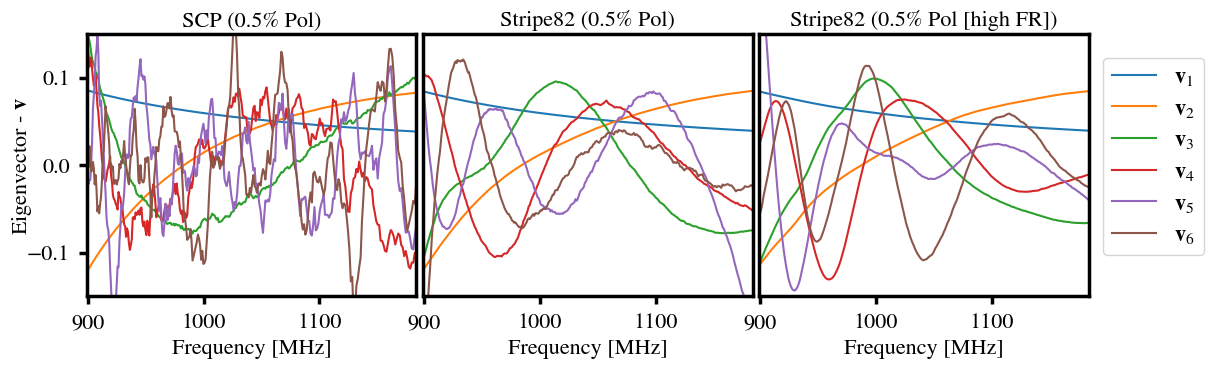}
    \caption{First six eigenvectors from the frequency-frequency covariance matrix for \oldred{both regions with 0.5\% polarisation leakage. \textit{Right}-panel is for the Stripe82 region for the high Faraday rotation case.}}
    \label{fig:Eigenvectors}
\end{figure*}

In \autoref{fig:ResidualFGs} the levels of foreground residuals for \oldred{both regions with 0.5\% polarisation leakage (and also the Stripe82 high-FR case)} are shown after PCA cleans with varying $\NFG$. These are calculated using the methods outlined in \secref{sec:FGresid}, which reconstruct the exact maps of the foreground-only signal remaining in the cleaned data (shown by solid lines in \autoref{fig:ResidualFGs}). For comparison, we also plot the \hinospace-only power spectra (dashed lines) calculated in a similar way by projecting the \hinospace-only simulated data along the $\NFG$ subtracted eigenvectors to precisely reconstruct the residual-\hi in the maps after the PCA clean.

\autoref{fig:ResidualFGs} shows that the residuals decrease with increasing $\NFG$ as expected, but the \hi signal is also damped. \oldred{The aim is to reach a level where the \hi signal dominates over the residual foregrounds, so they no longer bias results. It is encouraging to see that in general \hi dominates by an order of magnitude across small scales ($k\gtrsim 0.1\,\text{Mpc}/h$). In the Stripe82 high-FR case, a high $\NFG$ is required to bring the foreground residuals below the \hinospace-only power and even using $\NFG=14$, foreground residuals remain at a similar level to the \hi for modes with $k\lesssim 0.04\,\text{Mpc}/h$. We can see how results differ between the SCP and Stripe82 regions, for example in the SCP, a very mild clean ($\NFG=4$) is sufficient to achieve a residual foreground level an order of magnitude lower than the \hi across all scales, even in this 0.5\% polarisation leakage case.}

It is interesting to note how much the residual-\hi (dashed-lines) differ in \oldred{the high-FR case relative to the standard polarisation cases (e.g. comparing the \hi residual for $\NFG=16$) despite it being the same underlying simulated \hi data.} The reason for this is down to how the eigenvectors are constructed in each case and we show the first six in \autoref{fig:Eigenvectors}. For each \oldred{of the cases} we see little distinction between the first two eigenmodes, which are likely picking out the synchrotron slope and the free-free emission. But in the SCP the modes start to oscillate after this, which suggests that smaller scale cosmological \hi or noise is leaking into these modes. Whereas in the Stripe82 high-FR case, the eigenmodes are relatively smooth (except for the long wavelength oscillations caused by polarisation leakage \oldred{- which can be compared to the polarised spectra in \autoref{fig:FGSpectra}}) indicating that more large scale information will be removed. This is what we see in \autoref{fig:ResidualFGs} too, if we compare e.g. the $\NFG=16$ case for the two regions. In the \oldred{Stripe82 high-FR case} the small-$k$ is severely damped relative to the SCP, but comparing the scales around \oldred{$k=0.08 - 0.1\,h/\text{Mpc}$ we see that the Stripe82 high-FR $\NFG=16$ case} actually has slightly larger power in the residual \hinospace, owing to the fact that the eigenmodes being removed are smoother and contain less small-scale \hi power.

\autoref{fig:ResidualFGs} demonstrates the potential problem from additive biases caused by residual foregrounds, which will correlate in the auto-power spectrum. \oldred{We also see in \autoref{fig:2DPkPatch} and \autoref{fig:NFGcomparison} how these relatively small foreground residual levels can cause non-negligible additive biases in the total measured power spectrum.} Of course, in our simulated scenario we are able to decompose the contribution from residual foregrounds and \hi to the total power spectrum and make a well-informed choice on the optimal choice of $\NFG$. However, if dealing with real data, residual foregrounds and \hi would not be easily separable at the required precision. This highlights a central challenge to using \hi intensity mapping in auto-correlation for data that include foregrounds that can not be efficiently removed (see discussion in \secref{sec:FGresid}). 

\subsection{Cross-Correlation with Optical Galaxy Surveys}\label{sec:CrossCorrResults}

Previous \hi intensity mapping surveys with the GBT \citep{Masui:2012zc, Wolz:2015lwa,Wolz:2021ofa} and Parkes telescopes \citep{Anderson:2017ert,Li:2020pre} have relied on cross-correlations with optical surveys for successful detections of cosmological \hinospace. This is both due to  noise and systematic effects in these pathfinder intensity mapping experiments, but also due to foreground residuals. As we have already stated, intensity mapping simulations are typically idealised and foregrounds can be removed relatively straightforwardly. However, the real data in these early experiments have shown that to achieve low enough foreground residuals for successful detections, more aggressive foreground cleans are required compared to simulations. This is likely due to calibration issues and effects from polarisation leakage or chromatic beams, which can cause some frequency decoherence in the otherwise continuous foreground signals.

Cross-correlations with optical galaxy surveys are important because they allow for systematics to be mitigated (they drop out in cross-correlation), and a detection can be more easily achieved. We demonstrate this process with our simulated data sets and in doing so we can investigate the optimal level of foreground cleaning required. 

To do this we exclusively use our most dominant foreground region (the Galactic Plane), including polarisation leakage. This is an attempt to mimic real-data experiments, which as discussed often need high levels of foreground cleaning ($\sim10 - 20$ $\NFG$ modes removed from the data). As we have demonstrated in \autoref{fig:PkPatch} (third panel) and \autoref{fig:NFGcomparison}, the \hi auto-correlation is highly affected by the presence of such dominant foregrounds. Either dominant foreground residuals remain in the data from choosing $\NFG$ which is too low, or the largest scales are completely destroyed from choosing a higher $\NFG$.

\begin{figure}
    \centering
    \includegraphics[width=\columnwidth]{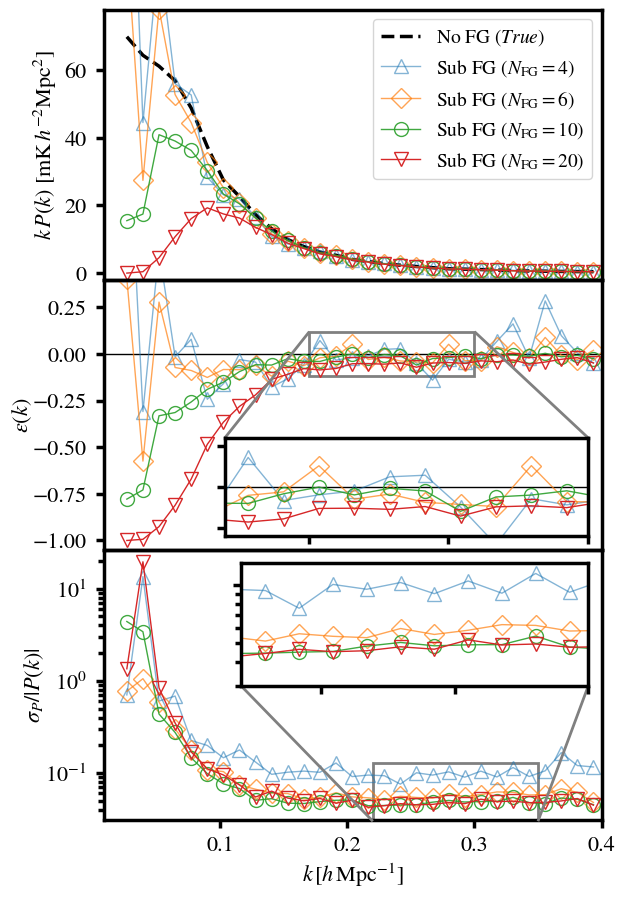}
    \caption{Results from the cross-correlation of optical galaxy data with \hi intensity maps cleaned using PCA with a range of $\NFG$. (\textit{Middle}-panel) shows $\varepsilon(k)$, the weighted difference defined by \autoref{eq:PkEps}, where positive (negative) values represent under- (over-)cleaning. We show the most foreground affected \oldred{case, the Stripe82 region with high-Faraday rotation polarisation leakage.} Comparison with previous results for this region, where an aggressive clean is needed and severely damps power in \hi auto-correlation, demonstrates how much cross-correlation improves results. \textit{Bottom}-panel shows that a less aggressive clean (lower $\NFG$) generally results in higher fractional errors $\sigma_P/P(k)$.}
    \label{fig:CrossCorrwithPol}
\end{figure}

\autoref{fig:CrossCorrwithPol} shows the improvements that can be made with cross-correlations. Here we cross-correlate the simulated optical data (outlined in \secref{OpticalSimsSec}) with the \hi intensity map data contaminated with a polarised Galactic Plane foreground and PCA cleaned. The top-panel immediately shows that a foreground clean with much fewer modes removed is sufficient for a cross-correlation measurement that has reasonable agreement with the no-foreground case. This is in spite of the large levels of residual foregrounds that will inevitably be remaining from such a mild clean. This is also shown in the middle panel where the weighted difference $\varepsilon$ (\autoref{eq:PkEps}) is plotted for each variant of $\NFG$. This shows that $\NFG=4$ and $\NFG=6$ deliver a reasonably consistent agreement across all scales with the foreground-free power spectrum. This also shows how more aggressive cleans still damp the power spectrum even in this case of cross-correlation. This is especially noticeable at large scales (small-$k$) but even at mid-range scales in the zoomed-in section at $0.17<k<0.3\,h/\text{Mpc}$, we can see power being damped as $\NFG$ is increased.

It is perhaps tempting to conclude that an optimal choice of $\NFG$ in cross-correlations can be entirely based on what delivers the best agreement with the foreground-free data, i.e. the $\varepsilon$ which is closest to zero. However, as is already slightly discernible from the middle-panel, a low choice of $\NFG$ does provide a higher variance in the results, as shown from the $\NFG=4$ case whose value for $\varepsilon$ fluctuates more than all other cases. This can be understood, by considering that in this situation a large amount of foreground residuals will be in the cleaned intensity map which cause more random and spurious correlations between the optical data and the foreground residuals. In other words, the higher levels of foreground residuals resulting from a lower $\NFG$ clean, will inevitably boost uncertainties in cross-correlations.

We investigated this further in the bottom panel of \autoref{fig:CrossCorrwithPol} which show the estimated fractional errors $\sigma_P/|P(k)|$ for each data point. We calculated these using the same approaches used in real-data studies (e.g. \citet{Masui:2012zc, Anderson:2017ert,Wolz:2021ofa}) by treating our \textsc{MultiDark} simulated data sets as real data. Alongside this we produced 100 lognormal mocks for both the \hi intensity maps and optical galaxy maps. From this we can calculate a foreground transfer function (see \secref{sec:TransferFunc}). Applying this transfer function to each of the mocks and then measuring the variance in their results provides an estimate for the power spectrum errors $\sigma_P$.

The bottom panel of \autoref{fig:CrossCorrwithPol} shows that errors are generally largest for the $\NFG=4$ case, highlighting the important point that foreground residuals will boost errors. At low-$k$ we also see the fractional errors are extremely high for the $\NFG=20$ case. This is because the power at these scales is damped so severely that the transfer function is essentially trying to recover a power spectrum from a position of $P(k)\sim 0$, which inevitably causes noisy results and boosts the variance. This demonstrates the balance required for an optimal choice of $\NFG$. An $\NFG$ which is too low causes too much foreground residual leading to large errors. An $\NFG$ which is too high, damps the power spectrum drastically causing too much scatter in the power recovered through the transfer function. There is also a requirement for an $\NFG$ balance at higher $k$-ranges. Looking at the zoomed-in section at $0.22<k<0.35\,h/\text{Mpc}$ we clearly see errors are highest for $\NFG=4$. They begin to decrease with an increasing $\NFG$ but we eventually find this saturates and there is little or no improvement in error even with a big jump from $\NFG=10$ to $\NFG=20$. This means going arbitrarily high in $\NFG$ will slowly stop improving errors but continue to bias results (increase $|\varepsilon|$).

The results from \autoref{fig:CrossCorrwithPol} are therefore strong evidence that a fine balance must be reached for an optimal choice of $\NFG$ in cross-correlations. This choice will also depend on the cosmological parameters being probed. For example, an investigation into primordial non-Gaussianity involves attempts to constrain the parameter $f_\text{NL}$, which requires large scales  (small-$k$) measurements. Going to $\NFG\sim 10$ in an attempt to minimise errors may not be plausible in this situation if the bias induced on these large scales is too strong. Conversely, probing something like the \hi abundance ($\Omega_\hinospace$), which generally just affects the amplitude scaling of the power spectrum and can thus be probed at most scales, could potentially allow for a more aggressive clean that controls errors but still does not heavily bias the higher-$k$ scales where this parameter can still be constrained. Therefore, it is unlikely that a universally optimal foreground treatment can ever be selected. This also supports conclusions from previous work that attempted to model the effects of foreground cleaning e.g. \citet{Cunnington:2020wdu} and \citet{Soares:2020zaq}. These works employed subtly different foreground modelling, which is likely due to the different range of scales they targeted given their science goals.

\section{Discussion}\label{sec:Conclusion}

Evidence from pathfinder 21cm intensity mapping data \citep{Masui:2012zc,Wolz:2015lwa,Anderson:2017ert,Wolz:2021ofa} suggests that we will, at least initially, be requiring fairly aggressive foreground cleans in \hi intensity mapping surveys. This is potentially due to instrumental responses to foregrounds causing effects such as polarisation leakage. Understanding the impact this has on probes of large-scale cosmic structure using \hi intensity maps is therefore paramount. In this paper we have provided a study into these issues by presenting a set of test data with a differing range of foreground contamination, both with and without effects from polarisation leakage. \oldred{We stress that the polarisation leakage model we use is far from being a precise emulation of these complex effects, nevertheless it captures the main issue with this systematic: its non-smooth and spatial-dependent behaviour in frequency, and a more advanced model is currently lacking.}

\oldred{Given our limited understanding of polarisation effects on intensity mapping, we have aimed to test extreme cases} where a high-$\NFG$ clean is required, which in this sense, is similar to early pathfinder experiments. This provides a means to begin investigating some of these issues in a simulated setting where we have full control and can separate contributions to the final observed signal. \oldred{We found that a varying polarisation leakage fraction does not create problems for a foreground clean (see \autoref{fig:2DPkPatch}) and an increasing fraction mainly just increases the amplitude of the eigenvalues one would remove anyway in a foreground clean (\autoref{fig:Eigenvals}). The main issue comes from the frequency decoherence to the foreground spectra which could plausibly be higher than predicted by the \texttt{CRIME} model we use. To investigate this further, we created a high Faraday rotation test using the polarisation leakage output from the Galactic plane region of \texttt{CRIME}. Whilst this mixing of regions does not represent a physically realistic scenario, it provides the necessary test case to investigate these problems with the models currently available.}

The contamination to the data from a foreground clean can manifest in two distinct ways: damping of cosmological \hi modes, and foreground residuals.\\
\newline
\noindent\textit{Damping  cosmological \hinospace}:

Inevitably, some cosmological \hi modes will be degenerate with the 21cm foregrounds and their information will be contained in the same modes that are being removed. This has the effect of damping the \hi power spectrum mostly on large scales (small-$k$). For our mild cases without polarisation leakage (and even the SCP with polarisation leakage), this effect is minimal (see \autoref{fig:PkPatch} and \autoref{fig:2DPkPatch}). However, where higher $\NFG$ cleans are needed, \oldred{as was required for our high Faraday rotation cases,} this damping becomes severe and generally isolated to small-$k_\parallel$ modes (see \autoref{fig:NFGcomparison}). In all cases though, a foreground transfer function (\secref{sec:TransferFunc}) should be able to compensate for this damping or alternatively, a foreground model with free nuisance parameters that can be marginalised over \citep{Soares:2020zaq}\oldred{, although neither approaches avoid the inevitable reduced sensitivity caused by the foreground clean}. \\
\newline
\noindent\textit{Foreground residuals}:

Another challenging effect from foreground contamination comes from residuals left in the data after a clean. In mild cases where foregrounds are removed relatively easily, foreground residuals are small but we have shown evidence that there can still be an additive bias (see high-$k$ values in \autoref{fig:FGremovalcomparison} with positive $\varepsilon$ values). However, previous work (e.g. \citet{Cunnington:2020mnn,Soares:2020zaq}) has shown that unbiased cosmological parameter estimates can be obtained even when contribution from residuals is not included, thus showing that for mild cases with low enough residuals, their effect is minimal. However, our results indicate that foreground residuals can be exacerbated by polarisation leakage. For example, \autoref{fig:ResidualFGs} demonstrated how even with large $\NFG$, the foreground residuals can still \oldred{have a similar amplitude to} the remaining \hi signal in some cases. This would cause additive biases and boost errors and needs to be modelled, as we outlined in \autoref{eq:PowerWithResidFG} and \autoref{eq:PowerErrorResidFG}. Quantifying the contribution from foreground residuals is not trivial with real data and this will be a key challenge for auto-correlation measurements. However, for cross-correlations with an overlapping optical galaxy survey the situation is somewhat simplified. Here the foreground residuals do not correlate with the foreground-free galaxy data, allowing less extreme cleans at the expense of a boost to errors (as demonstrated by \autoref{fig:CrossCorrwithPol}).
\\
\newline
\noindent Our investigation suggests that drawing general conclusions or recommendations for an optimal foreground treatment is not possible. An optimal method depends on the region being targeted, the instrumental calibration (e.g. susceptibility to polarisation leakage), and also on the scales being targeted depending on the survey's key science goals. But if one can achieve near-perfect instrument calibration, the results do appear fairly general. The \oldred{first column} of \autoref{fig:2DPkPatch} shows all regions can be cleaned by blindly removing 3 principal components and this delivers similar damping to \hi power and a similar level of residual bias. However, residuals may differ between regions for smaller-$k$ but without affecting the accuracy (measured by $\varepsilon$) and instead just affect the precision (boosting errors).

We introduced and tested three commonly employed blind foreground cleaning techniques; \fastica, GMCA and PCA, the latter being mathematically equivalent to SVD (\secref{sec:SVD}), and a polynomial fit. We found all three blind methods deliver essentially equivalent results in all cases. We discussed this in detail in \secref{sec:BlindComparison} -- in summary, this is due to \fastica\ and GMCA performing an initial PCA which they then try to improve upon by imposing spatial statistical independence and sparsity respectively. However, perhaps due to the sky size we use and the resolution, the foregrounds not included in the initial PCA reconstruction are not sufficiently sparse or non-Gaussian to be identified by GMCA or \fastica\ and no discernible improvement is made.

We also trialed a non-blind approach to foreground removal (\secref{sec:NonBlindGMCA}). Our tests revealed that this method can \oldred{potentially be competitive with a blind approach in the absence of polarisation leakage} (see \autoref{fig:FGremovalcomparison}). However, the method in its current form, is not robust to polarisation leakage effects and performs poorly when this is included. Thus, this approach would be reliant on a near-perfect calibration of the intensity mapping instrument. For these reasons, this approach would only likely be viable for future surveys where calibration strategies are highly optimised and astrophysical parameters can be tightly constrained, allowing more precise fits to the foregrounds. A potential benefit from this is the possibility of further combinations with a full-blind approach in a hybridization. \secref{sec:Hybrid} examined the cross-correlation between a non-blind cleaned map and a PCA cleaned one. This revealed that whilst little improvement can be gained in the accuracy of the final power spectrum, the foreground residuals in the two are subtly different and result in a lower contribution in the cross-measurement (\autoref{fig:HybridFGresids}). 

We hope our findings can be useful for analysing \hi intensity mapping data from the MeerKAT intensity mapping survey \citep{Santos:2017qgq, Li:2020bcr,Wang:2020lkn}, and for preparing cross-correlation strategies for MeerKAT and the SKA \citep{Carucci2017, Pourtsidou:2016dzn, Bacon:2018dui}.

\section*{Acknowledgements}
We are grateful to Paula Soares for helpful comments on the draft. SC is supported by STFC grant ST/S000437/1. JB, IPC and MI are supported by the European Union through the grant LENA (ERC StG no. 678282) within the H2020 Framework Program. AP is a UK Research and Innovation Future Leaders Fellow, grant MR/S016066/1, and also acknowledges support by STFC grant ST/S000437/1. This research utilised Queen Mary's Apocrita HPC facility, supported by QMUL Research-IT \href{http://doi.org/10.5281/zenodo.438045}{http://doi.org/10.5281/zenodo.438045}. We acknowledge the use of open source software \citep{scipy:2001,Hunter:2007,  mckinney-proc-scipy-2010, numpy:2011, Harris_2020}. Some of the results in this paper have been derived using the \texttt{healpy} and \texttt{HEALPix} package. We thank New Mexico State University (USA) and Instituto de Astrofisica de Andalucia CSIC (Spain) for hosting the Skies \& Universes site for cosmological simulation products. 

\section*{Data Availability}

The data underlying this article will be shared on reasonable request to the corresponding author.




\bibliographystyle{mnras}
\bibliography{Bib} 


\bsp	
\label{lastpage}
\end{document}